\documentclass[aps,pre,twocolumn]{revtex4} 

\usepackage{graphicx}
\usepackage{xcolor,color}
\usepackage{amsfonts,amsmath}
\usepackage{amssymb} 
\usepackage{appendix}
\usepackage{hyperref}
 \usepackage{ulem}
 \usepackage{upgreek}

\begin{document}

\title{
Nonlinear  redistribution  of wealth from a  stochastic approach}
 
\author{ Hugo Lima$^1$, Allan R. Vieira$^1$, Celia Anteneodo$^{1,2}$  } 
\address{$^1$Department of Physics, PUC-Rio, Rua Marqu\^es de S\~ao Vicente, 225, 22451-900, Rio de Janeiro, Brazil\\
$^2$National Institute of Science and Technology for Complex Systems (INCT-SC), Rio de Janeiro, Brazil}

\begin{abstract}
We investigate  the effect of applying nonlinear redistributive taxes to the yard-sale dynamics of assets. An amount of money is collected from each individual (tax) and   distributed back equally. 
We consider (i) a piecewise linear tax, 
exempting those with wealth below a threshold $w_0$, and taxing the excess wealth  otherwise, 
and (ii) a  power-law tax with exponent $\alpha>0$, which allows embracing regressive, proportional and progressive rules. 
The distribution of wealth obtained from numerical simulations of the agent-based dynamics is compared with the solution of its associated Fokker-Planck equation for  the  probability density function $P(w,t)$ of wealth $w$ at time $t$, in good agreement. 
Based on these solutions,  
we analyze how the different rules modify the distribution of wealth across the population, quantifying the level of inequality through the Gini coefficient. 
We note that the introduction of an exemption threshold does not always diminish inequality, depending on the implementation details. 
Moreover,  nonlinearity brings new stylized facts in the distribution of wealth compared to the linear case, 
e.g., negative skewness, bimodality, indicating stratification,  or a flat shape meaning equality populated wealth layers.  
\end{abstract}

\maketitle

\section{Introduction}

The tools of out-of-equilibrium statistical mechanics can be used
to investigate the dynamics of money~\cite{SM,kolkata,chakraborti2010,footnote}. 
The so-called random asset exchange models~\cite{kolkata,chakraborti2010},  
assuming wealth transfers  through pairwise interactions between agents,                       
have successfully shown the endogenous emergence of stylized facts of real wealth distributions such as the concentration of wealth in heavy tails and the formation of  a condensed layer of the poorest people~\cite{moukarzel2007,boghosian2019}.
In fact, rich-get-richer mechanisms can  lead to the continuous accumulation of wealth, accentuating inequality, a trend observed in most countries over time~\cite{richest,sciam,piketty}.  
Therefore, it is worth investigating possible mechanisms capable of counteracting or mitigating the different forms of wealth inequality.

By means of asset exchange models, diverse ways  to reduce inequality have been analyzed~\cite{Chakraborti2000,Chakrabarti2004,Mohanty2006,Iglesias2012,chorro2016,burda2019,sebas2020,benhur2020}. 
A direct way is to regulate the transactions, introducing asymmetries that favor the poorest people~\cite{Iglesias2012}. 
Savings, that is, 
when agents do not trade all their money on transactions but save a fraction, can also promote the reduction of inequality, under certain conditions~\cite{Chakraborti2000,Chakrabarti2004,Mohanty2006}. 
The impact of proportional taxes 
has also been studied~\cite{boghosian2014,boghosian2014a,boghosian2017,boghosian2018,boghosian2016}.
In this work, we will analyze the effect of nonlinear redistributive taxes, such that people pay taxes and receive subsidies, while performing random exchanges,  
without saving propensity.
The random exchanges are the  so-called yard-sale random transfers~\cite{hayes}, where individuals in an artificial society possess a certain wealth and participate in transactions by pairs, transferring a fraction of the wealth from one to another agent. 
The applied taxes can be regressive, progressive, or exempting the poorest population.

Besides the agent-based dynamics, we consider the Fokker-Planck description of the evolution of the probability 
density function (PDF) of wealth $P(w)$, 
an  approach  successfully applied before (see for instance, ~\cite{boghosian2014,boghosian2014a,burda2019}) to  extensions of the yard-sale dynamics.

The paper is organized as follows.  
The agent-based model and its Fokker-Planck counterpart are described in Secs.~\ref{sec:agent} and ~\ref{sec:FP}, respectively. 
In Sec.~\ref{sec:NL}, we present  results for the evolution of wealth distribution, and stationary states,  using the Gini index to   
compare how different tax protocols contribute to reduce wealth inequality. 
Final remarks are presented in Sec.~\ref{sec:final}.

\section{Agent-based description}
\label{sec:agent}

We start in a scenario without taxation, where $N$ individuals (or  agents)  are able to exchange money,  in a fully connected network, according to 
the yard-sale rule~\cite{ys1}. 
That is, two randomly chosen agents exchange a fraction $\epsilon$  of the smallest of their wealths, which are updated according to  the rule
\begin{eqnarray} \nonumber
w_i  &=& w_i + \epsilon \, \eta\,  min(w_i,w_j)\,,  \\
w_j  &=& w_j - \epsilon \, \eta\,  min(w_i,w_j)\,,
\label{eq:min_rule}
\end{eqnarray}
where $min (x,y)$ is the minimum between $x$ and $y$ and $\eta$ takes the values $\pm 1$ with equal probability (giving equal probability  for a trader to win or lose at each transaction)~\cite{boghosian2014}. As we will see below, $\epsilon$  can be related to the time scale of the dynamics. 
Although Eq.~(\ref{eq:min_rule}) describes an unbiased trading, it is well known 
that this dynamics leads the system to concentrations of wealth in the hands of few agents (oligarchy) and also the condensation of most agents below an extremely low level of wealth.

 While the yard-sale exchanges proceed, we apply a redistributive mechanism, in which agents pay a tax and the collected money is redistributed 
equally among all the $N$ agents. 

For an agent taxed at rate $R$, its wealth is updated as
\begin{eqnarray}
w_k  &=& w_k - R \, g(w_k).
\label{eq:agent1}
\end{eqnarray}
%
Since, the collected amount is equally distributed among all agents, then, 
\begin{equation}
  w_i = w_i + R\,  g(w_k)/N,\;\;\;\mbox{for all $i$}.
\label{eq:agent2}
\end{equation}

The unit of time is the Monte Carlo step (MCS),   corresponding to $N/2$ pairs of transactions given by Eq.~(\ref{eq:min_rule}).
The redistribution is applied at each period of time $\tau$.
That is, at each $\tau N/2$ iterations of Eq.~(\ref{eq:min_rule}), all the $N$ 
agents are taxed at the same time and the full collected  revenue  
is equally distributed among all agents. 
Therefore, the net losses and gains of  individual $i$, due to  this redistributive step, is  
\begin{equation} \label{eq:fw}
\Delta w_i =  R  \Bigl[  \sum_{j=1}^N g(w_j)/N - g(w_i) \Bigr]\equiv R f(\{w_j\}).
\end{equation}
Note that,   the total wealth $W$ is conserved in this model representing a closed economy. 
Although this assumption may be unrealistic depending on the time window considered, it allows a first approach to the problem.

\subsection*{Taxing kernels}
\label{sec:kernels}
   
\begin{figure}[b!]
	\centering	
	\includegraphics[width=0.23\textwidth]{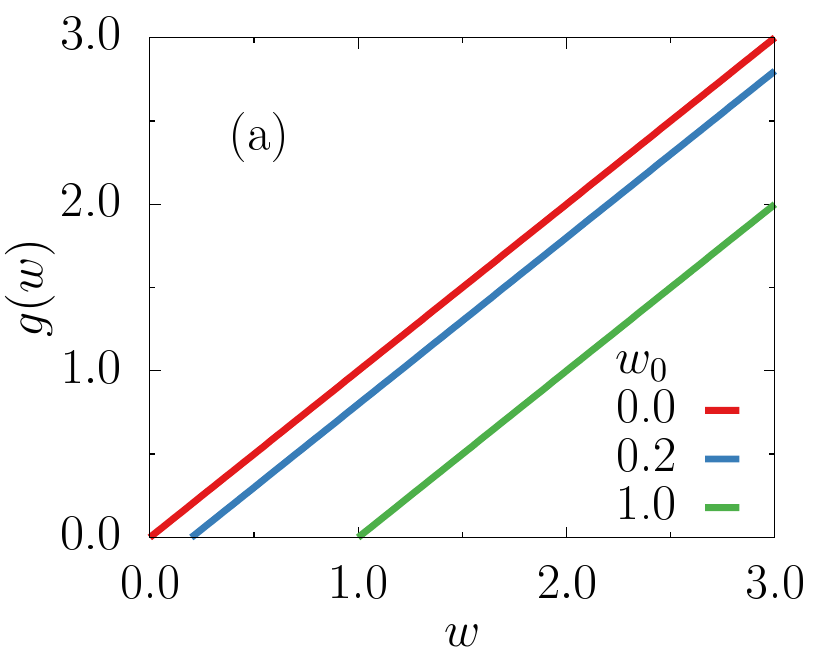} 
	\includegraphics[width=0.23\textwidth]{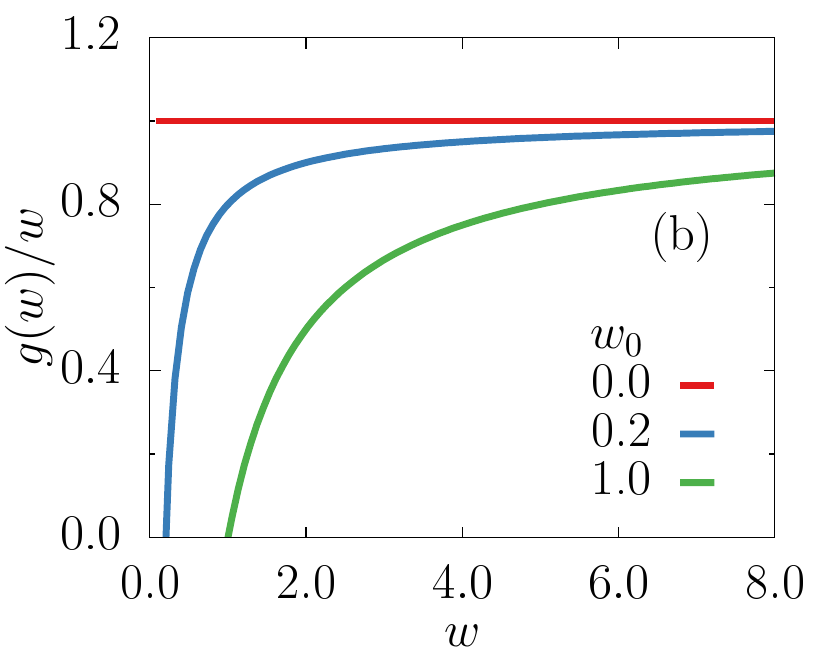}\\
 	\includegraphics[width=0.23\textwidth]{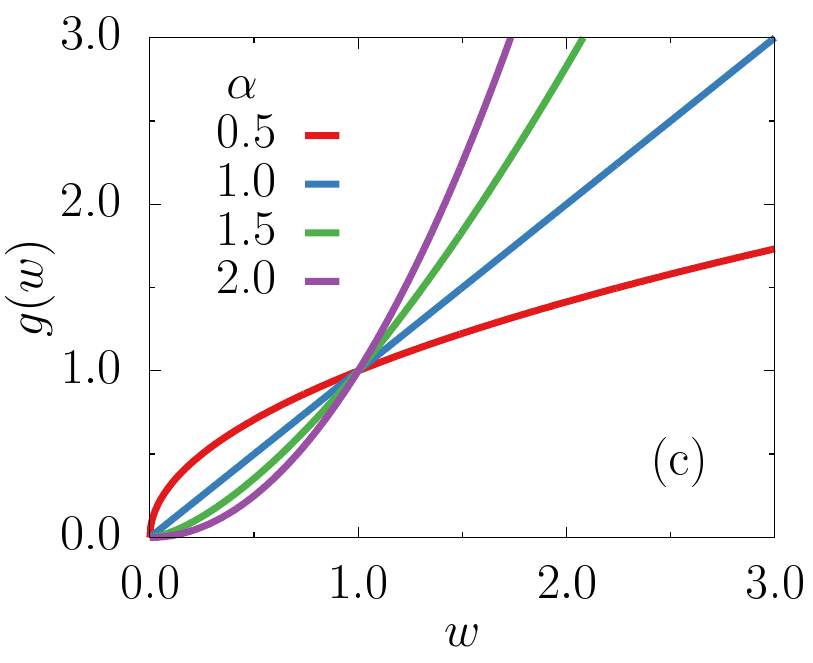}  
	\includegraphics[width=0.23\textwidth]{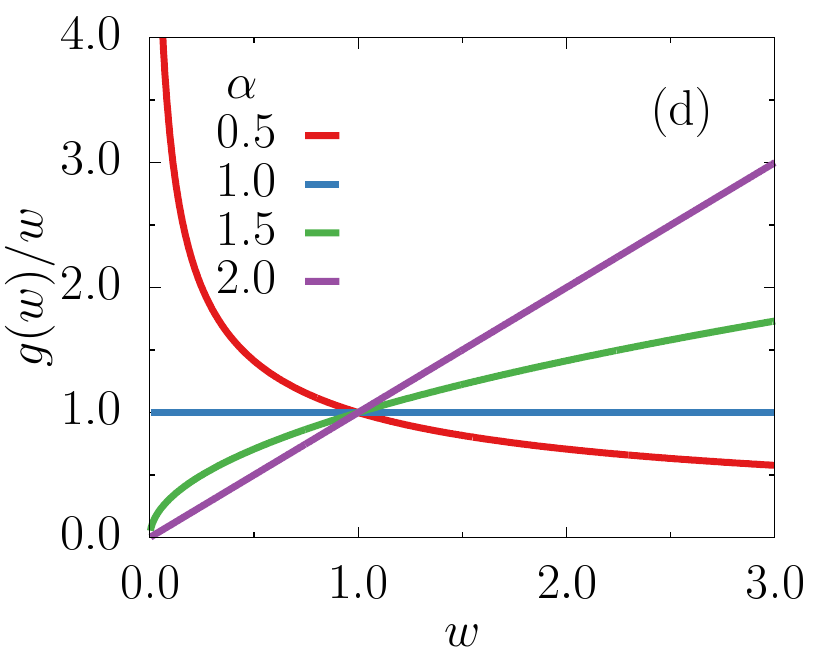} 
\caption{
	{\bf Taxing function $g(w)$ and tax rate $g(w)/w$},  for different values of 
the exemption threshold $w_0$	(a)-(b) , and of
	the power-law exponent $\alpha$ (c)-(d),  indicated in the legends. The linear case, recovered for  $w_0=0$ and $\alpha=1$, is also shown.
	}
	\label{fig:kernels}
\end{figure}

We analyze the effects of two different kernels that generalize the proportional case $g(w)=w$. 
One of them is the piecewise-linear function
\begin{equation} \label{eq:w0}
g(w) = (w-w_0) \Theta(w-w_0),  
\end{equation}
where $\Theta$ is the Heaviside step function and $w_0\ge 0$.
The proportional case is recovered by setting $w_0=0$.
Figure~\ref{fig:kernels}(a) depicts the 
 kernel (\ref{eq:w0}), for different values of $w_0$, and 
Fig.~\ref{fig:kernels}(b) shows the corresponding
 tax rate  $g(w)/w$, which is an increasing function 
 of $w$ for $w_0>0$, hence,  progressive.

We also consider  the power-law taxation 
\begin{equation} \label{eq:alfa}
g(w) = w^\alpha,
\end{equation}
with $\alpha>0$. 
This kernel shape is depicted in  
Fig.~\ref{fig:kernels}(c), and the 
corresponding  tax rate $g(w)/w$ in Fig.~\ref{fig:kernels}(d), where we note
that the taxation can be progressive (increasing tax rate, for $\alpha>1$), constant (if $\alpha=1$), or regressive (decreasing tax rate, for $\alpha<1$).

The nonlinear form of $g(w)$ implies different possible taxing schemes, which do not emerge in the linear case. We will consider the following two scenarios:

-   If $R$ is fixed, then the total collected money to be redistributed changes at each application of the redistribution scheme. Then, the subsidy, or gain per capita, $R \sum_{j=1}^N g(w_j)/N$ received by each individual  varies with time. 
This scenario is called setting "V".

- We also contemplate the scenario  "F" in which the subsidy is fixed,  kept constant at each application. Hence, the collected money needs to be adjusted. To accomplish this, the rate $R$   changes adaptively such that
$R \sum_{j=1}^N g(w_j)/N$ remains constant.


\section{Fokker-Planck description}
\label{sec:FP}
 
In the continuous-time limit, assuming small transactions, and uncorrelated Gaussian fluctuations,
the probability density function (PDF) of  the wealth $w$ at time $t$, $P(w,t)$, can be described by means of a Fokker-Planck equation (FPE). 
For linear taxation, the FPE  has been 
derived before by Boghosian~\cite{
boghosian2014, boghosian2014a}. 
The same FPE still holds in the current nonlinear case,  except that the form of the drift term is generalized. For selfcontainedness, let us sketch the steps. 
The FPE is given by~\cite{risken}
\begin{equation}
    \frac{\partial P}{\partial t}=
   \frac{1}{2}\frac{\partial^2}{\partial w^2}\left[D_2 P\right]
    -\frac{\partial}{\partial w}\left[D_1 P\right],
\end{equation}
where $D_1$ and $D_2$ are the Kramers-Moyal coefficients 
$D_j=  \lim_{\Delta t \to 0}  \langle (\Delta w)^j\rangle/{\Delta t}$, where the brackets average over the distribution of increments. Then,
\begin{eqnarray} \nonumber
D_1 &=&  \lim_{\Delta t \to 0} \frac{\langle (\Delta w)\rangle}{\Delta t} \simeq \frac{R}{\tau_0} f(w) \,,
\end{eqnarray} 
where, from  Eq.~(\ref{eq:fw}) in the continuous limit,
\begin{equation} \label{eq:fwcont}
  f(w)= \int_0^\infty g(w)\,P(w,t) \, dw   - g(w)\,,   
\end{equation}
and $\tau_0 = 2/N$ MCS is the minimal possible interval in simulations but 
$\tau_0 \to 0$ when $N \to \infty$. 
$D_1$ is  the deterministic drift associated to the redistributive action defined in Eq.~(\ref{eq:fw}).  
Notice that, in the linear case  $g(w)=w$, then   $f(w)=\bar{w}-w$, where $\bar{w}= \int_0^\infty w\,P(w,t) \, dw =W/N$.

The second Kramers-Moyal coefficient is~\cite{boghosian2014} 
\begin{eqnarray} \nonumber
D_2 &=&  \lim_{\Delta t \to 0} \frac{\langle (\Delta w)^2\rangle}{\Delta t}
\\ \nonumber
&\simeq&   \frac{\epsilon^2}{\tau_0} \int_0^\infty dw' P(w'){[w\Theta(w'-w)+w'\Theta(w-w')]}^2\\ \nonumber
&=&   \frac{\epsilon^2}{\tau_0} 
\underbrace{\int_w^\infty dw' P(w')}_{\displaystyle A(w)} w^2 +  \frac{\epsilon^2}{\tau_0}  \underbrace{\int_0^w dw' P(w')w'^2}_{\displaystyle B(w)} \\
&=& \frac{\epsilon^2}{\tau_0}  \left[ w^2A(w)+B(w)\right]\,,\label{eq:M2}
\end{eqnarray}
emerging from  the stochastic contribution of the yard-sale transaction at each iteration  (in a time interval $\tau_0$), $  \Delta(w,w',\eta)  = \epsilon\, \eta\, [w\,\Theta(w'-w)+w'\Theta(w-w')]$. 
Notice that $A(w)$ is the complementary cumulative distribution and $B(w)$ is the incomplete second moment, with the normalization condition
$\int_0^\infty dw' P(w') =1$.

Finally, scaling time  as
$t\epsilon^2/\tau_0 \to  t$,   defining  $\chi=R/\epsilon^2$,  in the limit of large $N$, we can finally write  
\begin{equation} \label{eq:FPE}
\frac{\partial P}{\partial t} = \frac{1}{2}\frac{\partial^2 }{\partial w^2} 
\biggl(  \Bigl[  w^2 A +B \Bigr]P    \biggr)   
-  \chi \frac{\partial }{\partial w} 
\bigl( f P    \bigr)   \,.
\end{equation}

The first term in the right-hand side of Eq.~(\ref{eq:FPE}) can be interpreted as   
a state-dependent diffusive spreading of wealth, which has its origin in the microscopic random exchanges between agents. In the second term, the drift $\chi f(w)$ represents the net gain (loss) suffered by agents with wealth $w$, due to the redistributive tax. 
 If $\chi=0$, the PDF evolves 
accentuating the condensation 
at $w=0$ and the  heavy Pareto power-law tail at large $w$~\cite{boghosian2014}, respectively meaning increase of the population at extreme poverty and concentration of wealth among a small number of people (oligarchy). 
This endless process that accentuates inequality can be broken when $\chi>0$ by the flow of wealth  from the richest to the poorest people.

Let us finally formulate explicitly the form of the drift $\chi f(w)$ in each setting presented at the end of Sec.~\ref{sec:kernels}.

In setting V, the gain due to redistribution is variable. Using Eq.~(\ref{eq:fwcont}), the drift can be written as 
\begin{equation} \label{eq:f1}
\chi f_{V}(w)=  \chi[\beta(t)\bar{w}  - g(w)]  ,
\end{equation}
where   
\begin{equation}
\beta(t) =\int_0^\infty g(w)\,P(w,t) \,dw/\bar{w} .
\end{equation}  

In setting F,  the subsidy (gain per capita) is fixed, then we can write 
\begin{equation} \label{eq:f2}
\chi{f}_{F}(w)= \chi[ \bar{w} - \gamma(t) g(w)], 
\end{equation}
where, from Eq.~(\ref{eq:fwcont}), $\gamma$ must be adjusted according to
\begin{equation} \label{eq:gamma}
 \gamma (t)= \bar{w}/\int_0^\infty g(w)\,P(w,t) \,dw \,.
\end{equation} 
In the linear version, where $g(w)=w$~\cite{boghosian2014,boghosian2017}, we have $\beta=\gamma=1$, 
meaning time-independent taxes, hence, a unique setting.

We characterize the PDFs of wealth  in terms 
of a widely used inequality indicator,  the Gini coefficient, which can be estimated as~\cite{boghosian2017},

\begin{equation}  \label{eq:gini}
G(t)=1-\frac{2}{\bar{w}}\int_0^\infty   x P(x,t)A(x,t) dx \,.
\end{equation}

In order to follow the time evolution of the wealth PDF, we numerically solved   Eq.~(\ref{eq:FPE}) though a standard forward-time centered-space scheme (details are shown in Appendix~\ref{app:time}). 
Typically, we used as initial condition, 
one resulting from the driftless evolution, at a time where the  Gini index is $G \simeq 0.59$.

Alternatively, to obtain the steady state solution, we  directly solved the 
stationary FPE, that is, setting the time derivative 
in Eq.~(\ref{eq:FPE}) equal to zero, as described in Appendix~\ref{app:steady}. 
Indeed, this solution coincides with the long-time solution of the FPE. 
In all numerical examples, we set $\bar{w}=1$. 

%

\section{Comparison between agent-based and FPE descriptions}
\label{sec:match}

\begin{figure}[b!]
	\centering	
	\includegraphics[width=0.45\textwidth]{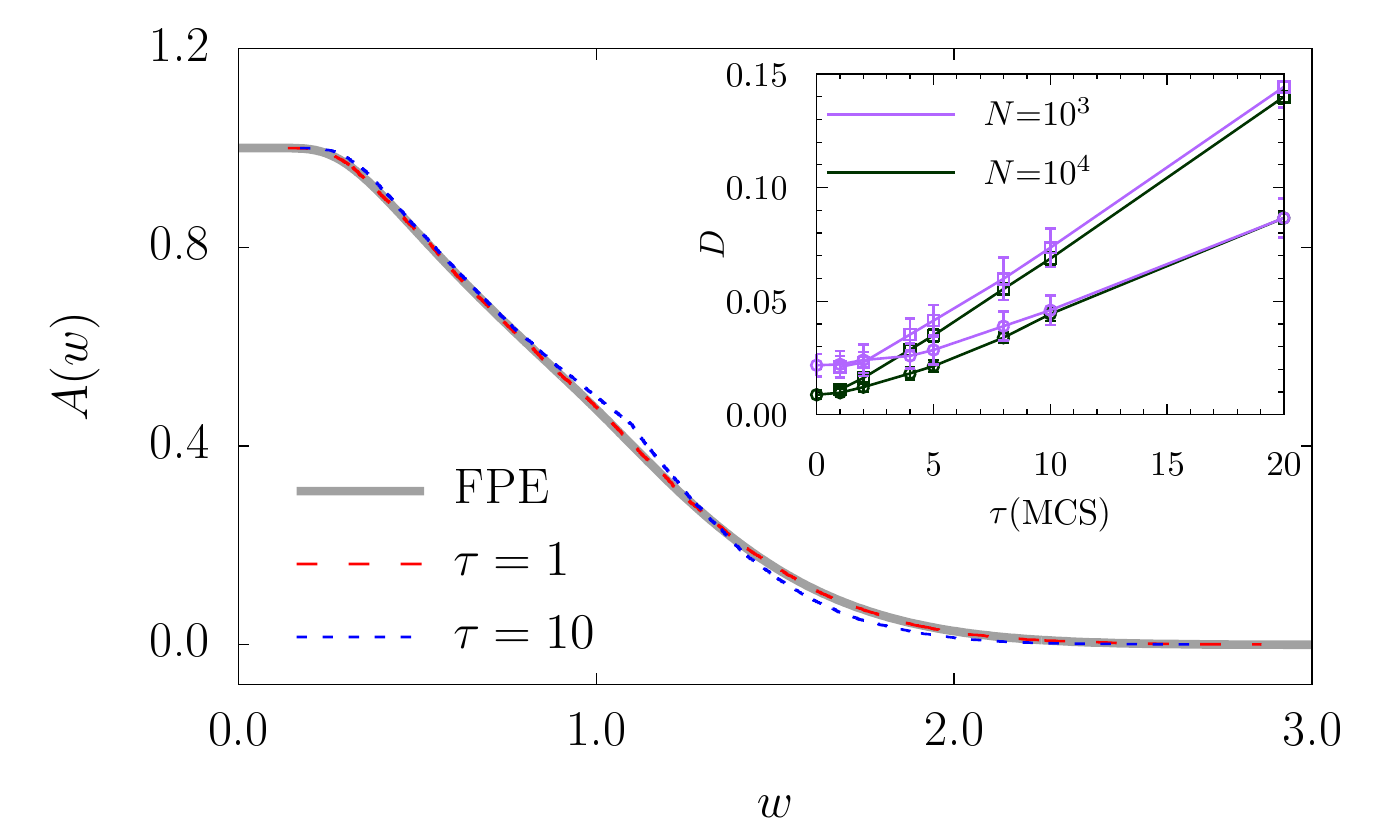}  %
\caption{
	Comparison between the cumulative PDFs obtained from agent-based simulations (colored dashed lines)
	and  Fokker-Planck solutions (solid gray line), 
	for different values of the redistribution period $\tau$,   at the instant $t=40$ MCS. We chose the piece-wise linear kernel in Eq.~(\ref{eq:w0}) with $w_0 = 1.0$, and $\chi /\tau= 1.0$, within scenario F (fixed subsidies). In simulations, we considered $N = 10^4$ agents and   $\epsilon = 0.1$ in Eq.~(\ref{eq:min_rule}). In the inset we depict the  KS distance $D$ as a function of $\tau$ for two different values of $N$. 
	The distances immediately before and after redistribution are measured in each case. Symbols and error bars correspond to the average and standard deviation over 100 samples. 
	}
	\label{fig:comparison}
\end{figure}

In Fig.~\ref{fig:comparison}, we provide a 
comparison  between the distributions obtained form numerical simulations of the agent-based model and from the numerical integration of the FPE. 
The cumulative PDF is considered, instead of the PDF, in order to smooth the fluctuations of numerical histrograms. 
The parameter $\tau$, which gives the time elapsed  between successive taxations, is varied. 
The agreement is very good for the minimal period $\tau=2/N$, even when $N=10^4$, and becomes better for increasing $N$ as expected.
To quantify discrepancies, we measured the Kolmogorov-Smirnov (KS) statistic $D$ between cumulative distributions. 
In the inset of  Fig.~\ref{fig:comparison}, we compare 
the  distributions obtained from agent-based simulations and from the solutions of the FPE, plotting $D$ as a function of the period $\tau$ of redistribution. 
First, notice that the distance between simulations and theory for $\tau=2/N$ (first point in the figure), decreases with $N$, as expected since
the KM coefficients of the FPE are defined for $\Delta t \to 0$ (and $\tau\to 0$).
We observe that the distance $D$ increases with $\tau$, however, the solutions of the FPE still provide a good prediction for larger $\tau$, at least for the unequal initial condition considered. 
 
\begin{figure}[h!]
	\centering	
	\includegraphics[width=0.45\textwidth]{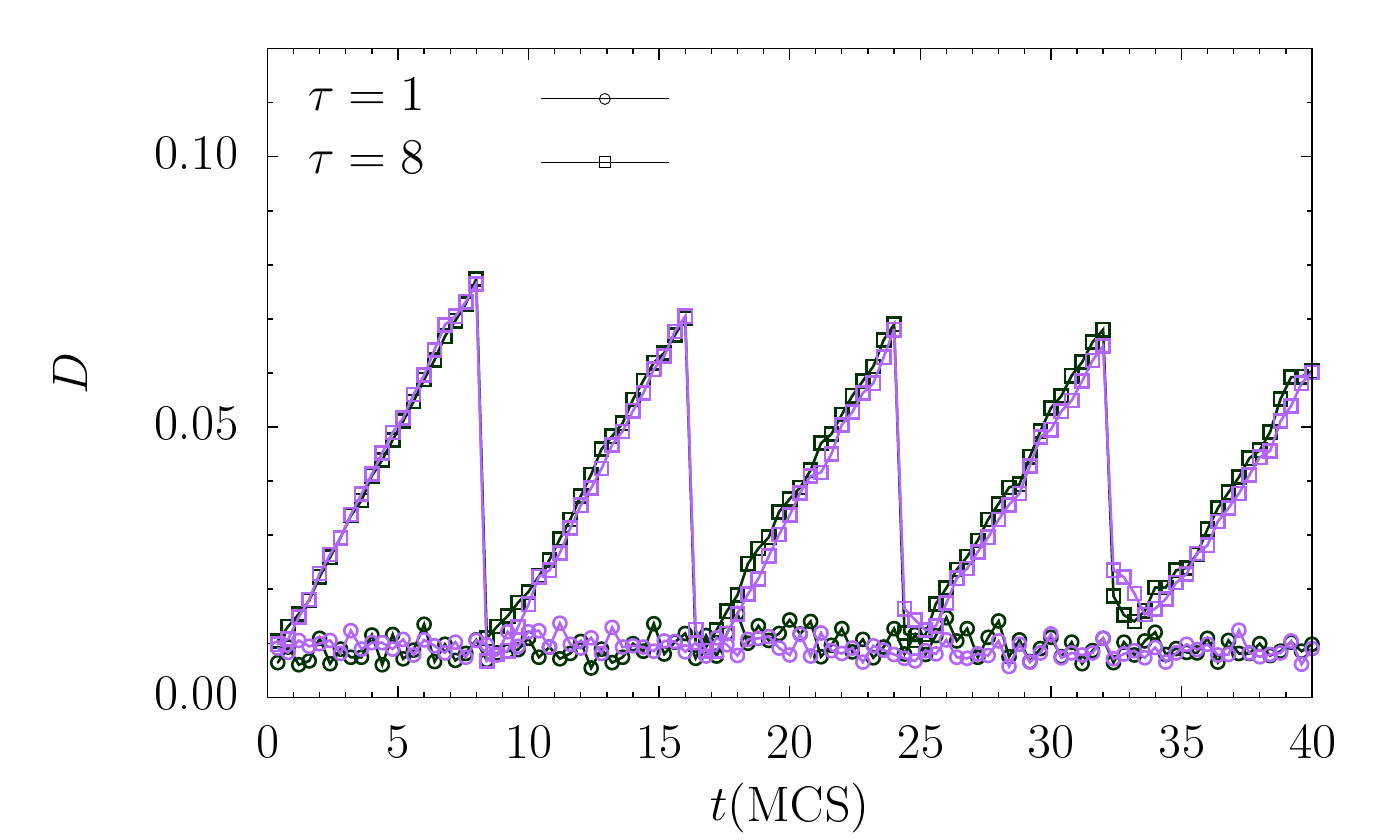}   
\caption{
KS distance vs. time $t$. At each  instant, we compute the   distance between the PDF for a given value of $\tau$ (indicated in the legend) and the PDF for the smallest period $\tau=2/N$. 
For each $\tau$, two different samples are shown. 
	The remaining parameters are the same in Fig.~\ref{fig:comparison}. For $\tau=1$ the distance is within noise level. For larger $\tau$, oscillations become noticeable, when only diffusive spreading occurs $D$ increases, but the application of the distributive step makes the distance minimal.  
	}
	\label{fig:comparisonxt}
\end{figure}

In Fig.~\ref{fig:comparisonxt}, we compare  with the case $\tau=2/N$, the cumulative PDFs obtained for $\tau=1$ (circles) and $\tau=8$ (squares), 
through  the KS statistics vs. time $t$.
During the period between taxes, only the diffusive spreading acts and the distance increases, when the redistribution is performed, the distance decays to the lowest level. Therefore, if $\tau$ is larger than the typical scale of diffusion, then a steady is not reached but rather an oscillatory regimes settles. 

Let us note that, for comparisons, 
we used  constant $\chi/ \tau$ ($ =1$), since the taxation rate must increase 
if the period increases,  to maintain the effective taxing rate invariant. 
In order to compare time scales, and to convert the scale $t$ of the FPE into the simulation time, we had to revert the 
transformation done in Sec.~\ref{sec:FP}, setting $t  \to t \tau_0/\epsilon^2$. 
Let us also note that, excessively large $\chi$ could lead to negative wealth (see Appendix~\ref{app:bounds}), but this was not observed in the studied cases.

Given that the FPE provides a good prediction of agent-based results, even for not too small $\tau$,  and since   the solutions of the FPE  provide smoother curves than
histograms from simulations, then we will show 
in what follows results based on FPE solutions only.

\section{PDF evolution and steady state solution}
\label{sec:NL}

As shown in previous literature~\cite{boghosian2014}, in the absence of regulation ($\chi=0$), Eq.~(\ref{eq:FPE}) leads to progressive 
condensation towards $w=0$, and a heavy tail for large $w$ develops, without actually reaching a normalizable steady state.
The inclusion of the drift with $\chi >0$  allows attaining a steady solution, 
in all the cases considered.

For each family of kernels $g(w)$, we   develop in Secs.~\ref{sec:w0} and \ref{sec:alfa}, the setting F (fixed returns): adjusting $\gamma$ selfconsistently to produce the same quantity returned (or subsidy) to all. 
Setting V  will be 
discussed later in Sec.~\ref{sec:variable}.

\begin{figure}[b!]
	\centering	
	\includegraphics[width=0.35\textwidth]{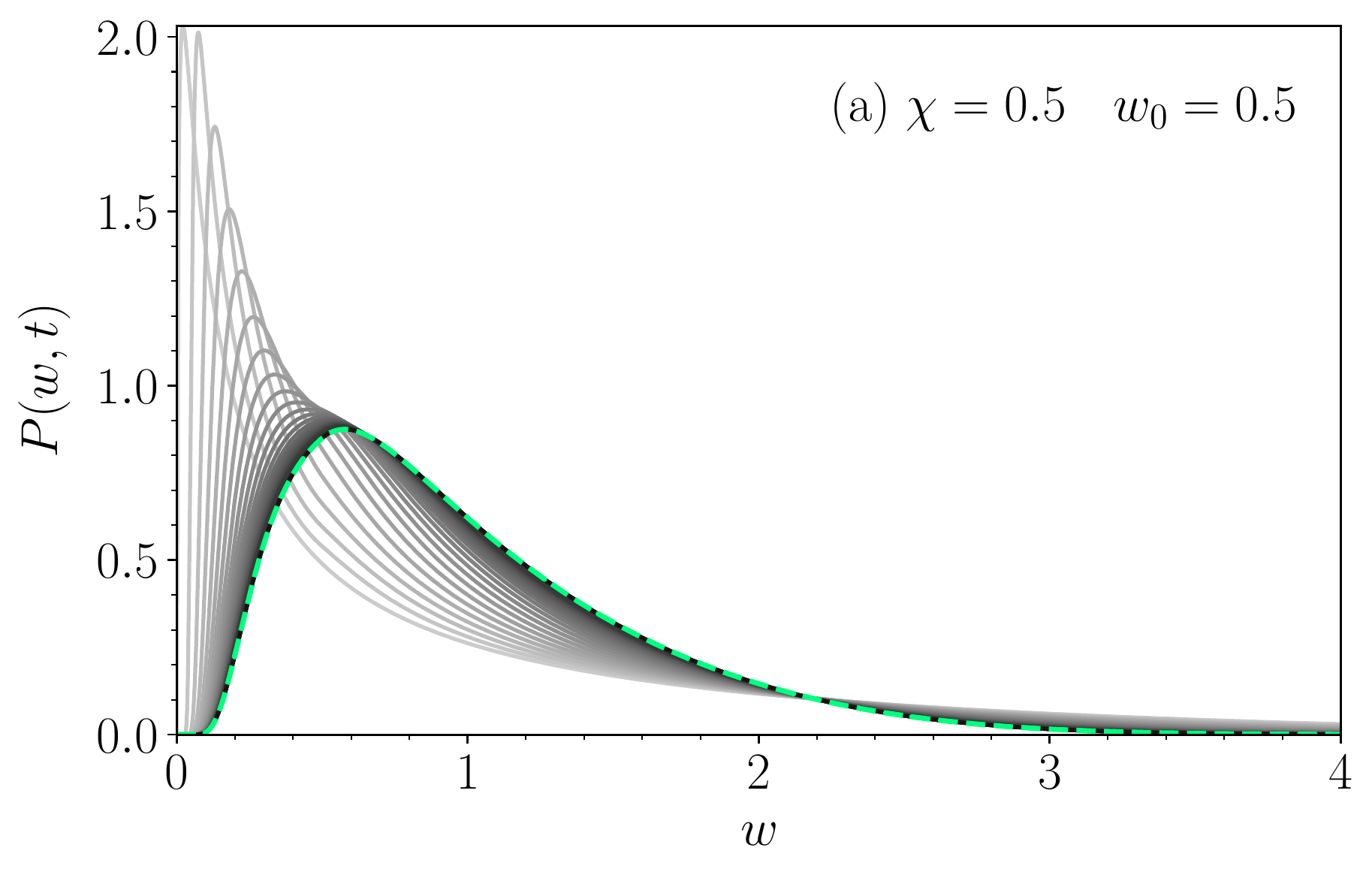}
	\includegraphics[width=0.35\textwidth]{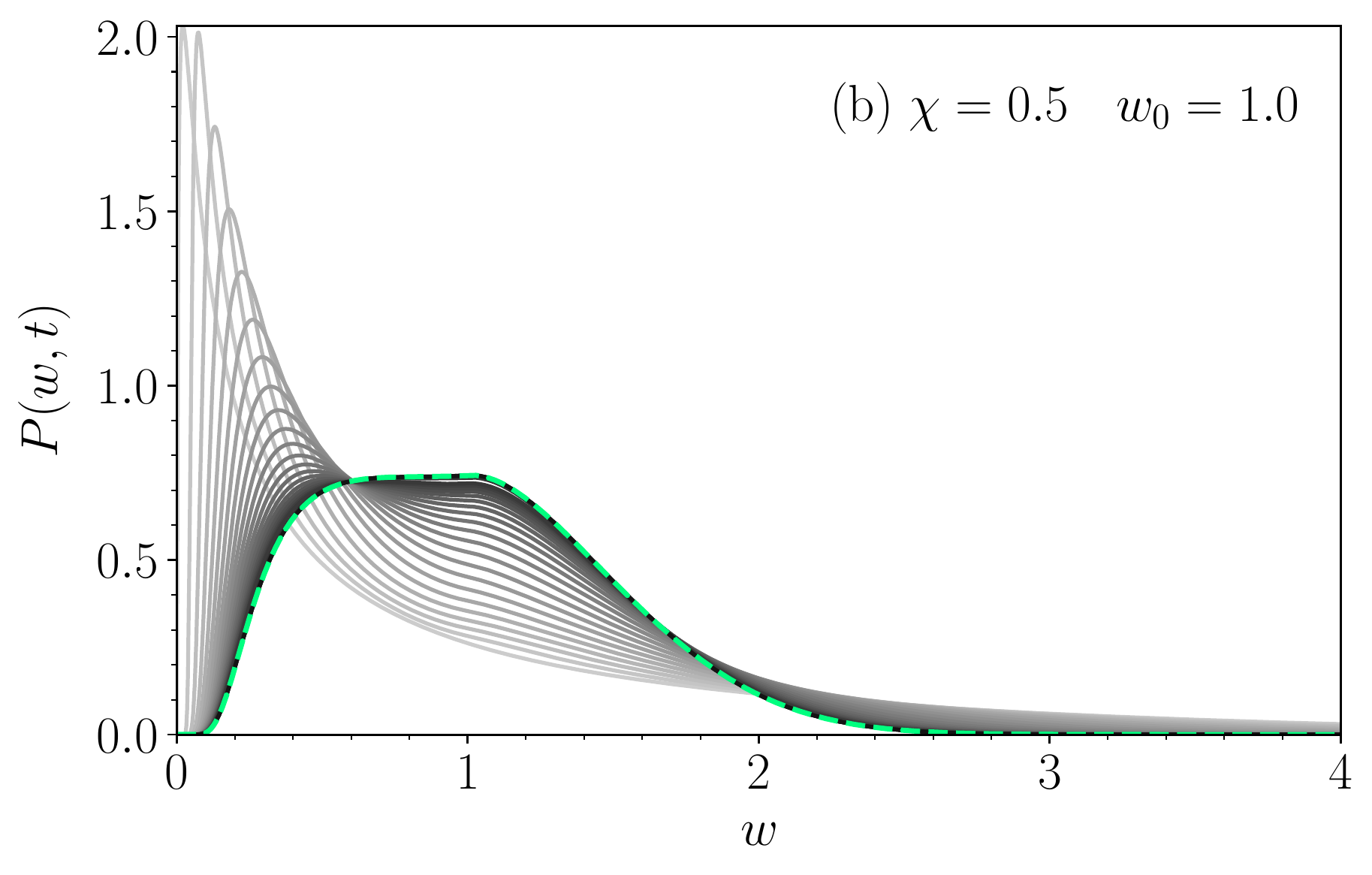}
	\includegraphics[width=0.35\textwidth]{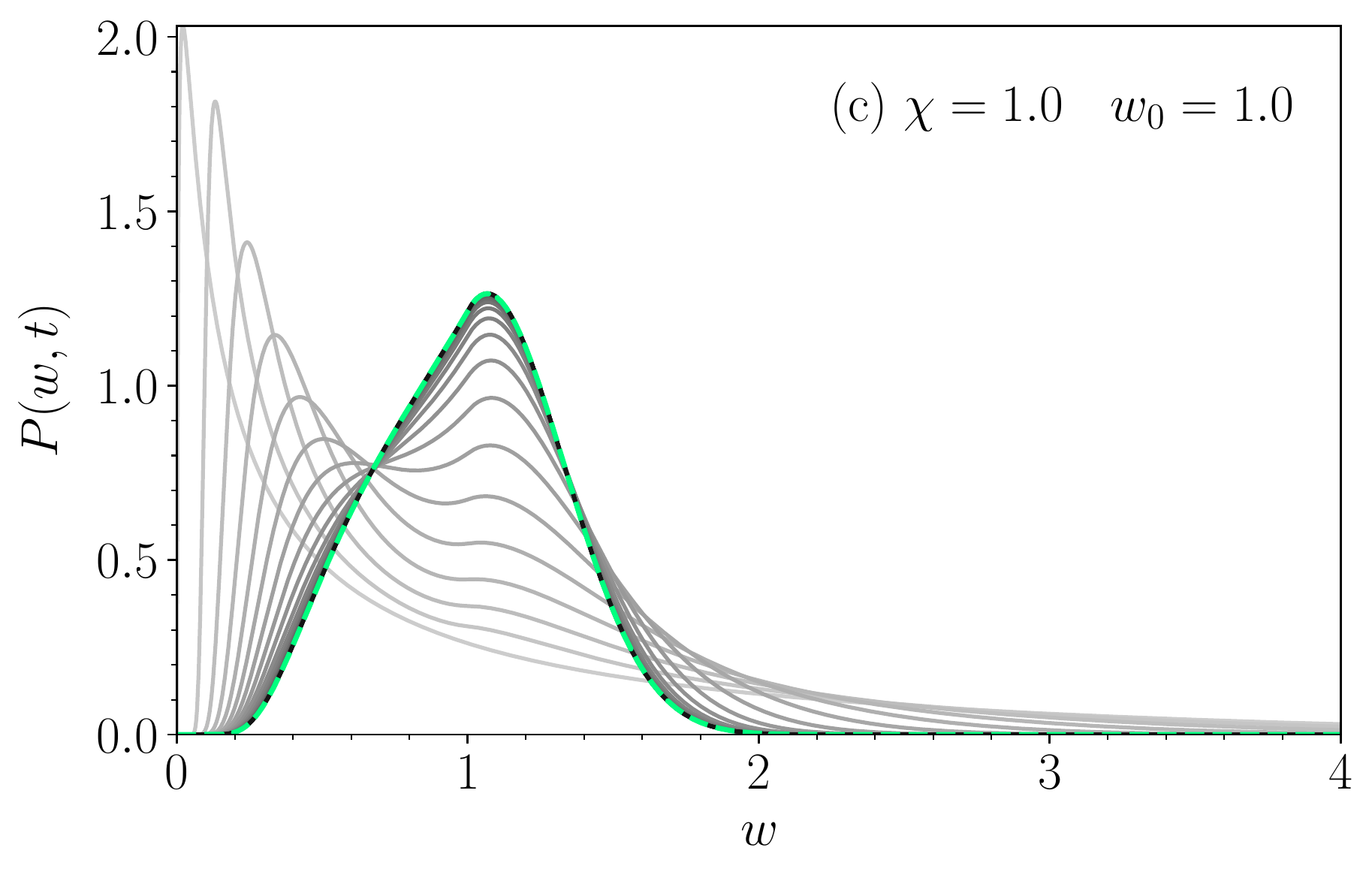}
\caption{
	{\bf Evolution of the wealth PDF with piecewise-linear drift}, 
 for values of $\chi$ and $w_0$ indicated in the legends, 
	at times increasing (from lighter to darker)
	from $t=0$   to $t=2$ at each $\Delta t=0.1$, 
	and from $t=2$ to $t=5$  at each $\Delta t=0.5$.
    The	solution of the stationary FPE is also plotted (green dashed line). 
	}
	\label{fig:PDFw0_t}
\end{figure}

\subsection{Piecewise-linear tax with  exemption limit $w_0$}
\label{sec:w0}

The evolution of $P(w,t)$, under the taxation ruled by the piecewise linear kernel defined in Eq.~(\ref{eq:w0}),  is shown in Fig.~\ref{fig:PDFw0_t} for different values of $\chi$ and $w_0$, 
starting from an initial condition that results from the driftless evolution, at a time where the  Gini index is $G \simeq 0.59$. 
In all cases, increasing $\chi$ makes the final state less spread. 
The condensation at $w=0$ is suppressed, meaning the absence of a majority at extreme poverty, and  the heavy tail for large $w$ becomes restricted by a cutoff, reducing large fortunes and oligarchy.
In Fig.~\ref{fig:PDFw0_t}a, for $w_0=0.5$, we find a picture qualitatively similar to that of the linear case~\cite{boghosian2014}, where the PDF has a positive skewness. 
Raising the threshold $w_0$, the PDF can become almost flat 
(see Fig.~\ref{fig:PDFw0_t}b), meaning uniformly populated wealth layers. 
For even larger $w_0$, the skewness is  inverted at long times, with 
a mode larger than the average value $\bar{w}=1$ (e.g., Fig.~\ref{fig:PDFw0_t}c). 
These are new features introduced by the piecewise function. The evolution of the Gini index and further details are shown in Appendix~\ref{app:time}.

\begin{figure}[b!]
 \centering
	\includegraphics[width=0.35\textwidth]{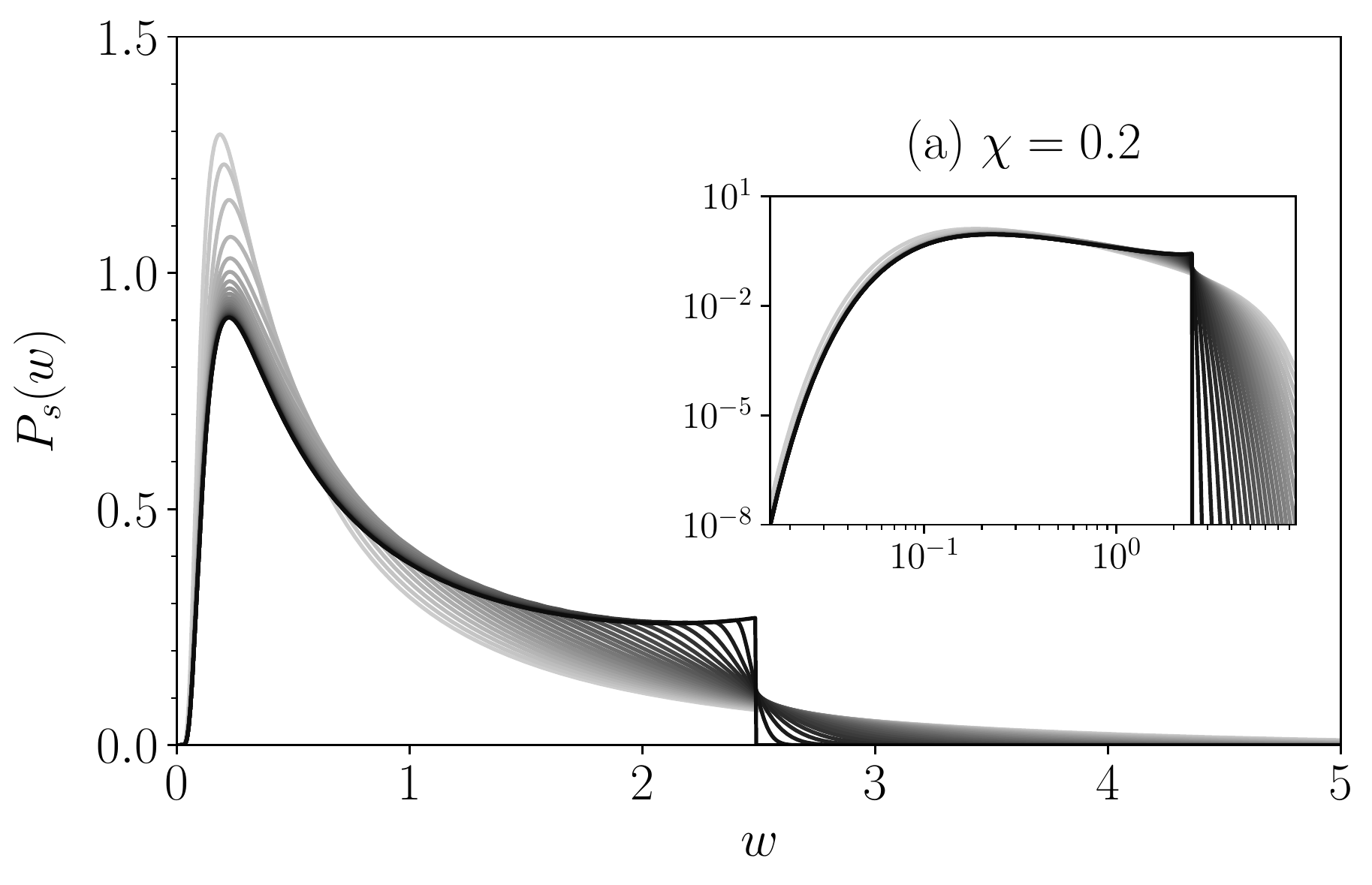}
	\includegraphics[width=0.35\textwidth]{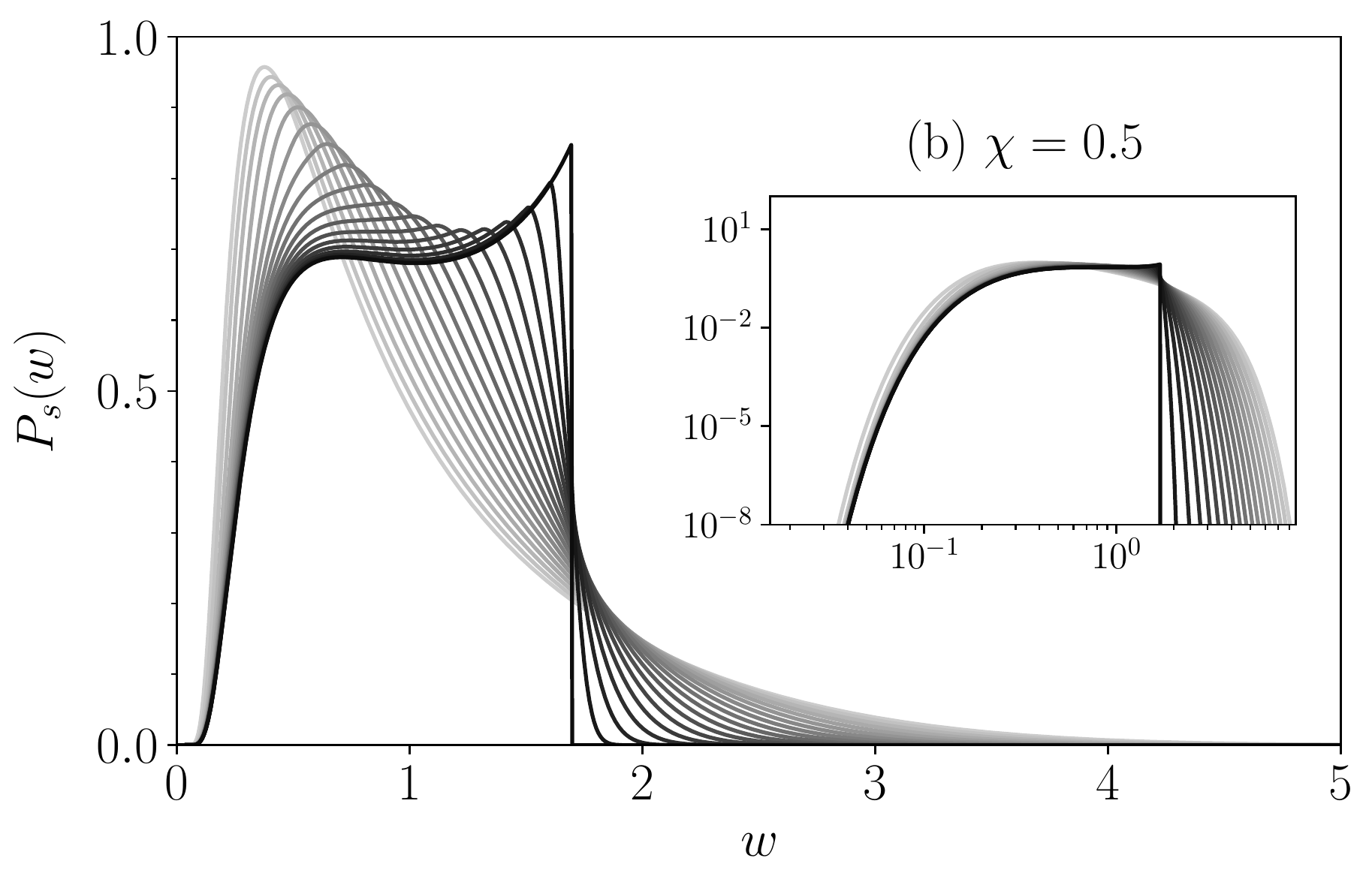}
	\includegraphics[width=0.35\textwidth]{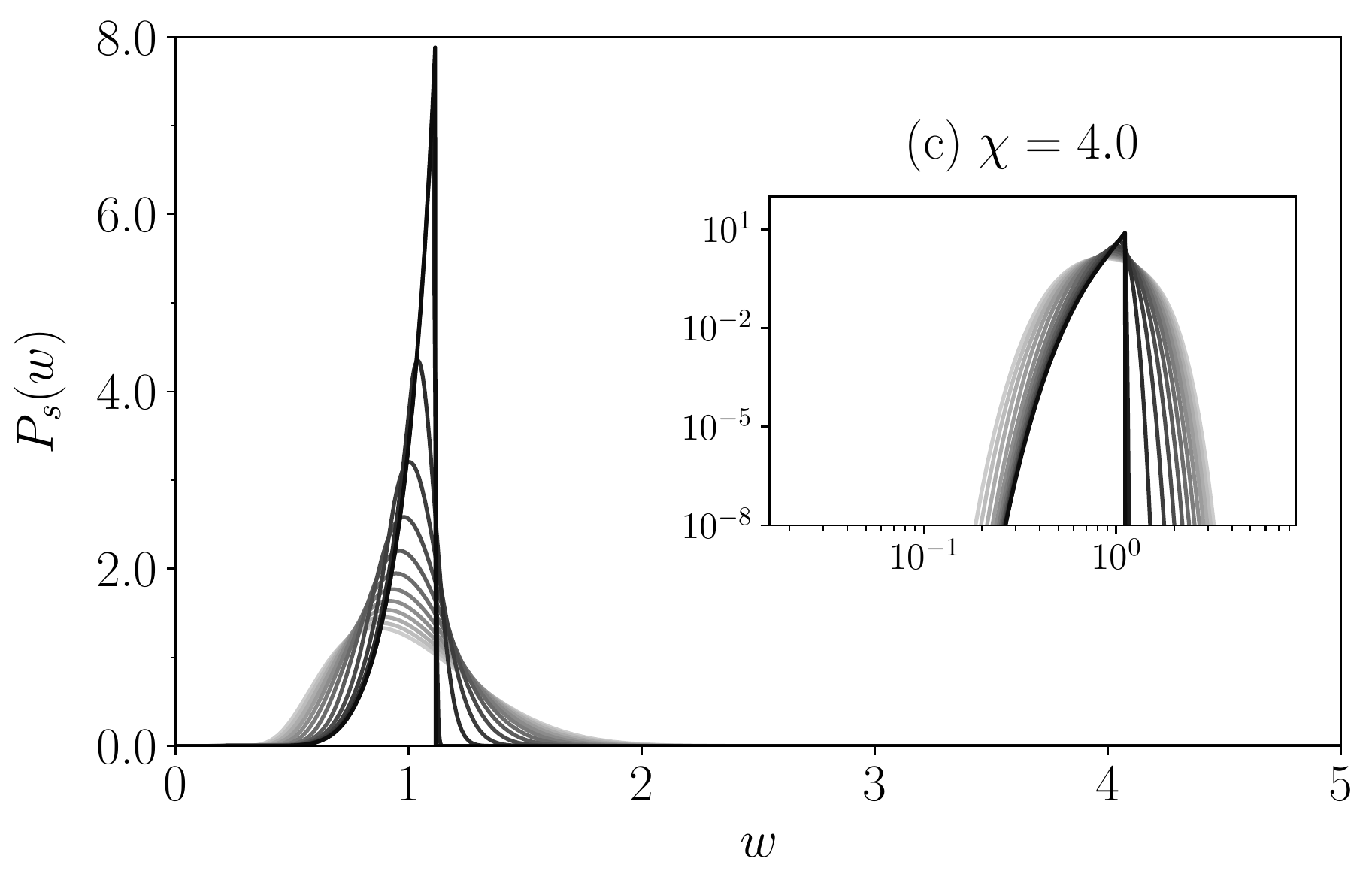}
\caption{
	{\bf Stationary wealth PDF, with piecewise-linear drift}, for fixed $\chi$ according to the legend, and different values of $w_0$ 
 varying (from light to dark lines) at each  
	$\Delta w_0=0.1$ from 0.0 up to the limiting value 
	$w_0 \simeq $2.5 (a), 1.7 (b) and 1.1 (c). 
	Inset: same data in log-log scale to exhibit the cut-offs. 
	}
	\label{fig:PDFw0_s}
\end{figure}

The stationary PDFs for fixed  
$\chi$ and a large set of values of $w_0$
are shown in Fig.~\ref{fig:PDFw0_s}. 
Increasing the rate $\chi$ narrows the PDF around the 
mean, particularly, the  cut-off at low $w$ is shifted such that wider ranges of poverty are eliminated.  This behavior is also observed in the linear case and is not noticeably affected by $w_0$~\cite{boghosian2017}.
The exemption threshold $w_0$ affects more strongly  the intermediate and large-$w$ layers. 
The PDF can become bimodal (e.g, Fig.~\ref{fig:PDFw0_s}b),  
indicating classes with defined asset level, but scales are 
not well separated.  
The change of skeweness can be observed in these plots for 
not too small $\chi$ (Figs.~\ref{fig:PDFw0_s}b-c). 
Moreover, in general, raising the threshold $w_0$ leads to 
 a more effective cut-off such that assets that surpass that level tend to be suppressed. 
 This is due to the flow of  wealth above $w_0$   
towards the population with lower assets, depopulating the large-$w$ tail.

 \begin{figure}[h]
 \centering
	\includegraphics[width=0.35\textwidth]{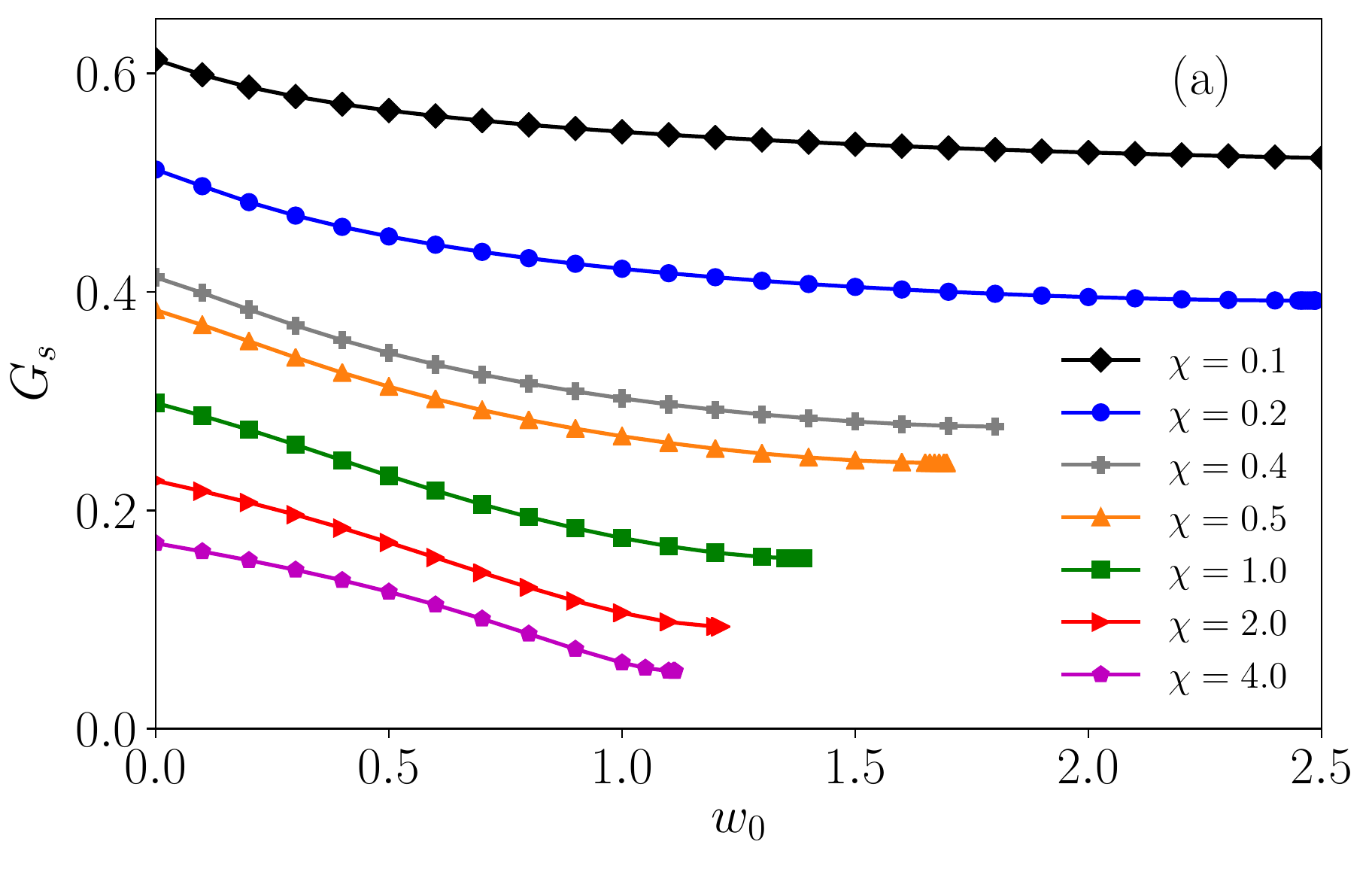}
	\includegraphics[width=0.35\textwidth]{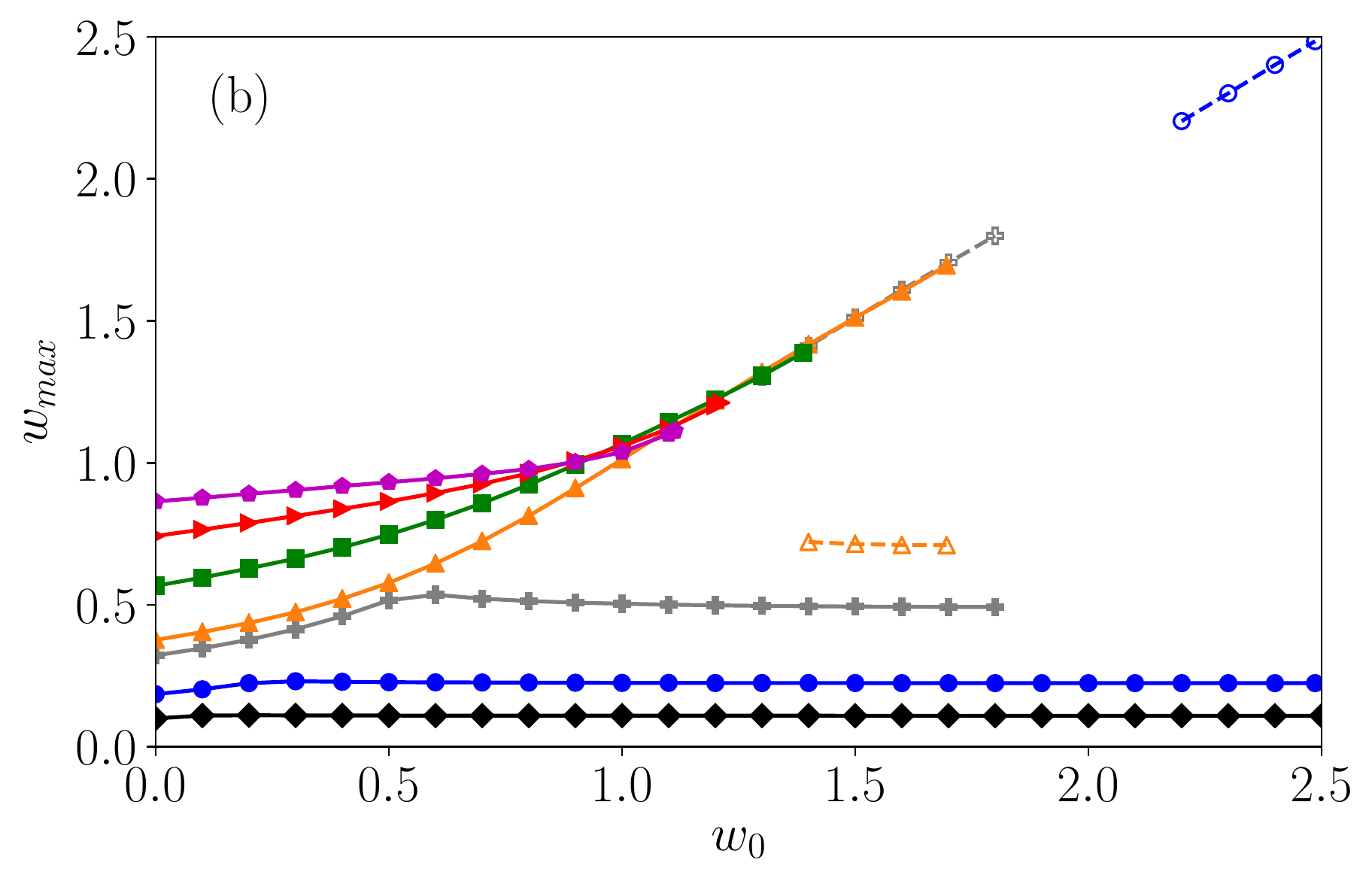}
	\caption{
	(a) {\bf Stationary  Gini index} for different values of $\chi$ as a function of $w_0$. (b)  {\bf Mode of the wealth PDF} (filled symbols)  vs. $w_0$. The second maximum, when it exists, is also plotted (hollow symbols). Lines are a guide to the eye. 
	}
	\label{fig:giniw0_s}
\end{figure}

The   long-time value $G_s$ of the Gini index is shown for different values of $\chi$ in Fig.~\ref{fig:giniw0_s}a. 
It decreases by raising $w_0$, as expected, 
however, a finite minimal level is attained   
at a limiting value of $w_0$. 
The larger is $\chi$, the more sensitive is $G_s$ to $w_0$. 
We have considered the full range of values for completeness, 
although some intervals of the parameters may be unrealistic.   
For instance, at the limiting value of $w_0$, the slope $\gamma_s$ becomes divergent, to keep the average $\bar{w}$ fixed (see Appendix~\ref{app:time}),
and the PDF becomes truncated as can be observed in Fig.~\ref{fig:PDFw0_s}. 

In Fig.~\ref{fig:giniw0_s}b, we plot the mode, $w_{max}$ (filled symbols), as well as the second maximum (hollow symbols) whenever it exists. 
For low values of $\chi$, the mode is weakly sensitive to $w_0$ and remains below 
$\bar{w}$ (e.g., case $\chi=0.2$), producing 
positively skewed distributions.
For larger $\chi$, the mode is shifted towards the line $w_{max}=w_0$, and can exceed the 
mean value $\bar{w}$ (unity, in our examples). 
When a second peak at larger $w$ develops 
(which occurs for not too large $\chi$), 
it can become the mode as $w_0$ increases (e.g., for   $\chi=0.5$ in Fig.~\ref{fig:giniw0_s}b, and also see Fig.~\ref{fig:PDFw0_s}b).


\subsection{Power-law taxation}
\label{sec:alfa}

\begin{figure}[h!]
	\centering	
	\includegraphics[width=0.35\textwidth]{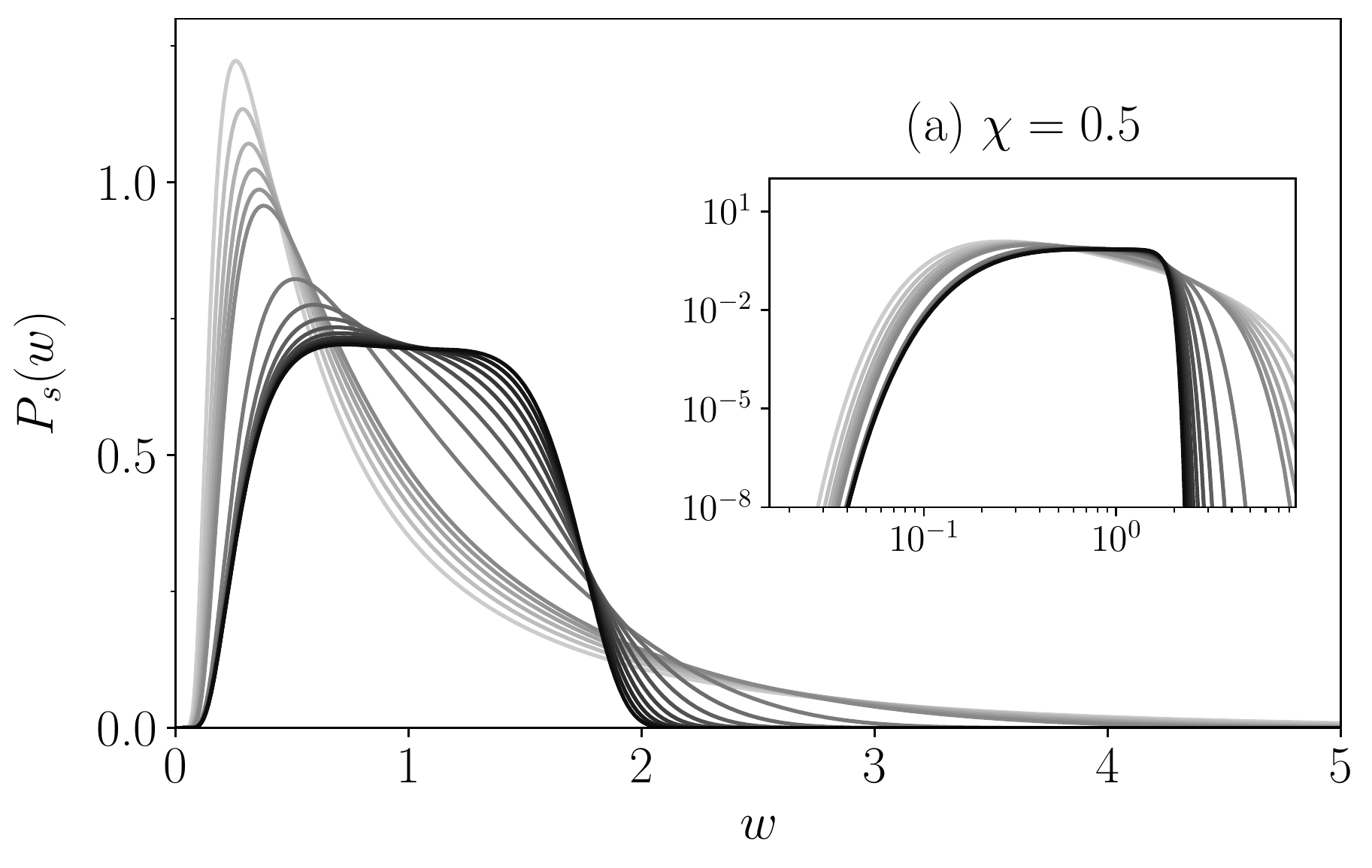}
    \includegraphics[width=0.35\textwidth]{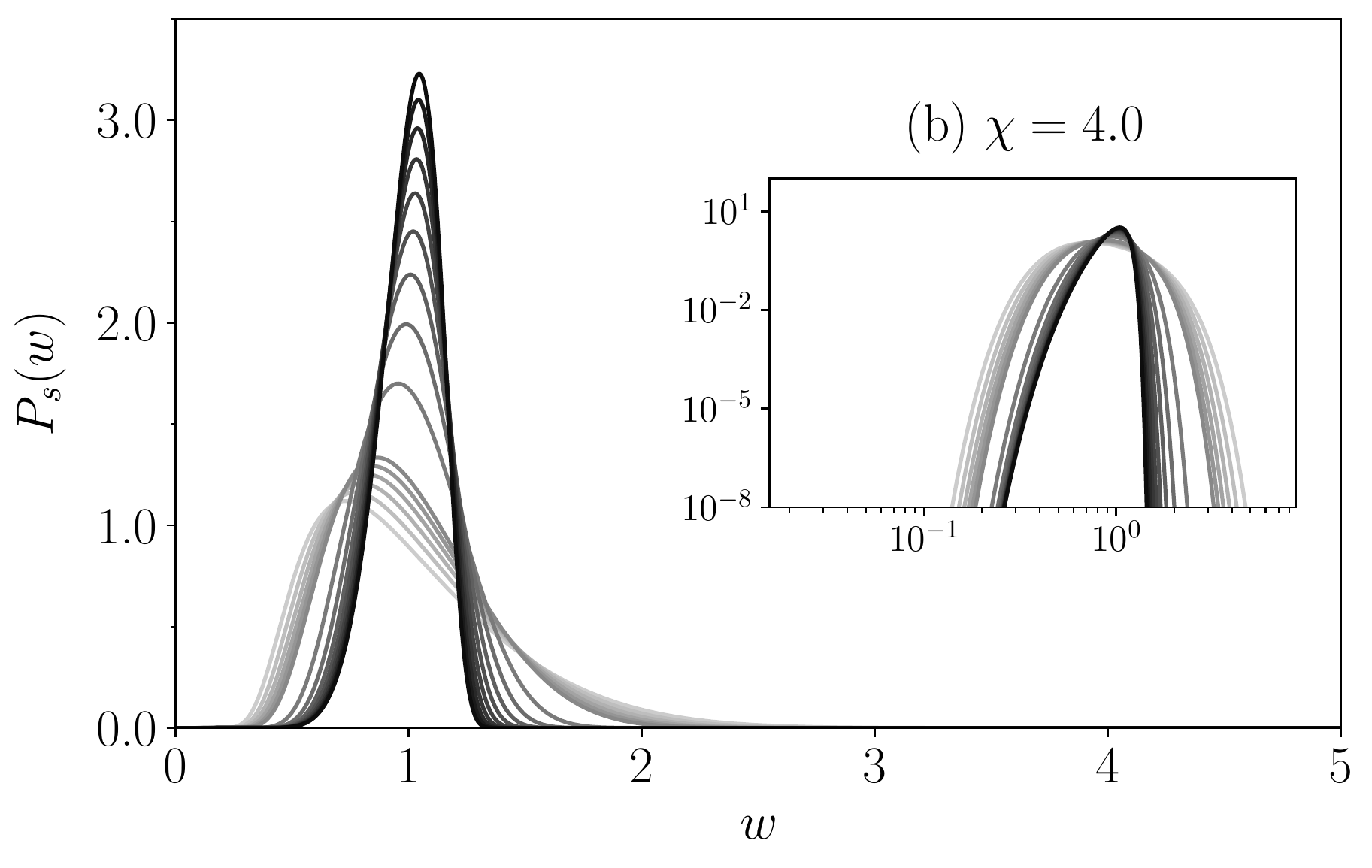}
\caption{
	{\bf Stationary wealth PDF, for power-law drift}, $P_s(w)$, for different values of $\alpha$ 
	at each $\Delta \alpha =0.1$ for $\alpha \in [0.5,0.9]$, and 
	$\Delta \alpha =1$ for $\alpha \in [1,10]$, with values of $\chi$ according to the legend. 
	Inset: same data in log-log scale to show the cutoff. 
	}
	\label{fig:PDFalfa_s}
\end{figure}

Stationary PDFs for different values of $\alpha$ and  $\chi$ are 
exhibited in Fig.~\ref{fig:PDFalfa_s}. 
As in the piecewise-linear case, there is a narrowing of the PDF with increasing $\chi$. It is noteworthy that similar effect is observed when the saving propensity increases~\cite{Chakraborti2000}.
When $\alpha>1$,
flat, bimodal, and negatively skewed PDFs can also emerge as for the piecewise-linear kernel,  but the PDF shapes are smoother for the power-law kernel and  the cutoffs  less sharp.

\begin{figure}[t!]
 	\centering	
	\includegraphics[width=0.35\textwidth]{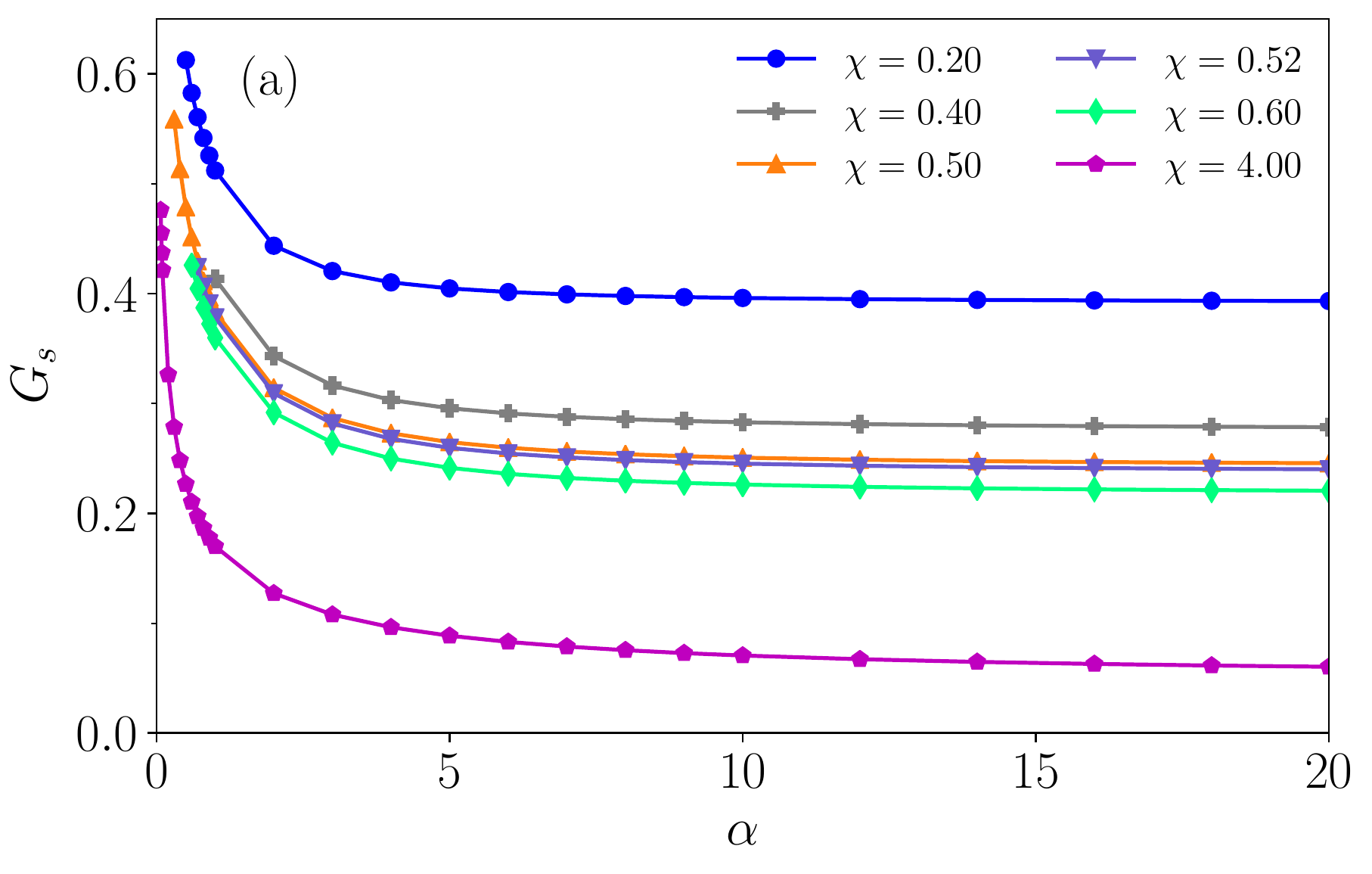}
	\includegraphics[width=0.35\textwidth]{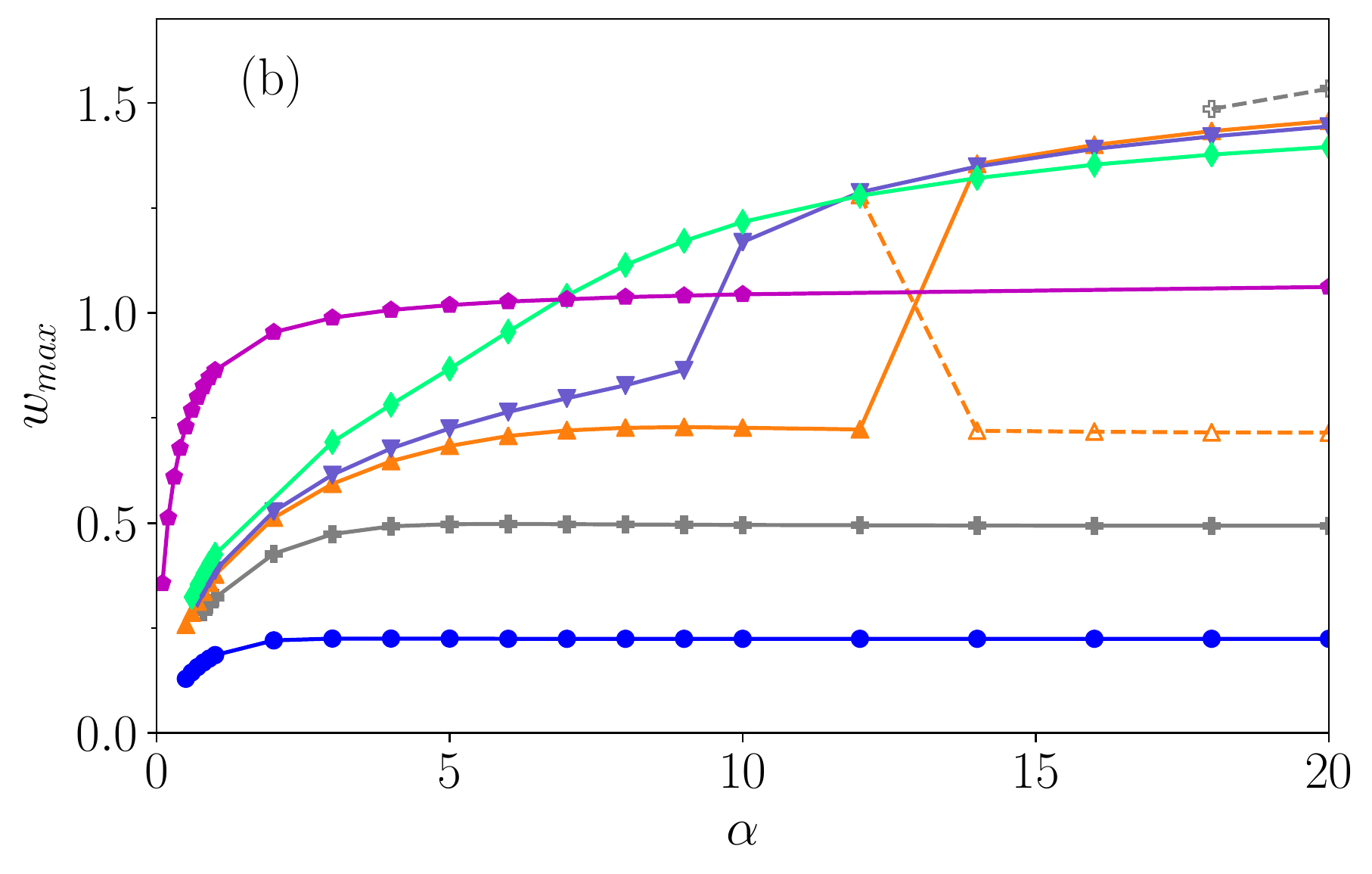}
	\caption{
	(a) {\bf Stationary Gini index} for different values of $\chi$ as a function of $\alpha$. (b) {\bf Mode of the stationary wealth PDF} (filled symbols) vs. $\alpha$. The second maximum, when it exists, is also plotted (hollow symbols). Full lines are a guide to the eye.
}
	\label{fig:ginialfa_s}
\end{figure}

The stationary values of index $G$ vs. $\alpha$, 
for fixed values of $\chi$, 
are presented in Fig.~\ref{fig:ginialfa_s}a. 
When comparing these curves with the respective ones of the piecewise-linear case in Fig.~\ref{fig:giniw0_s}a, 
we notice a matching for  $\alpha \simeq 4 w_0 \gtrsim 1$. 
While the progressive taxation with $\alpha>1$ emulates better the tax with exemption threshold $w_0$, 
we find the main differences for $\alpha<1$, due to its regressive character. 
In the limit $\alpha \to 0$,  
 from Eq.~(\ref{eq:gamma}), we have 
 constant $\gamma=\bar{w}$.
Then
\begin{equation} \label{eq:a0}
f(w)= (\bar{w}-\gamma)=0  \,, 
\end{equation}
implying that the FPE drift term goes to zero in that limit. 
Therefore, the dynamics 
evolves towards condensation and long tails.   
Even though, the sublinear kernel  produces relatively 
low values of the Gini index for large enough $\chi$. 
In fact, $G_s$ rapidly decreases with $\alpha$ in the regressive case. 
With regard to the maxima shown in Fig.~\ref{fig:ginialfa_s}, 
the mode overcomes the average $\bar{w}$ and a second maximum 
can appear only for very large values of the exponent $\alpha$.

\subsection{Setting V (variable returns)}	
\label{sec:variable} 	

Up to now, we worked out setting F, 
fixing the amount of wealth shared 
and adapting the coefficient $\gamma(t)$, consistently to conserve the average wealth.
Now, we analyze setting V, where  $\beta(t)$ is adjusted. 
The stationary solutions can be related
by identifying each term of the drift   in Eqs.~(\ref{eq:f1}) and (\ref{eq:f2}), 
yielding
\begin{eqnarray}
   \beta_s   &=&  1/\gamma_s 
   \nonumber\\
      \chi_{V} &=&\chi_{F}\gamma_s.  \label{eq:corresp}
\end{eqnarray}

\begin{figure}[t!]
	\centering	
	\includegraphics[width=0.35\textwidth]{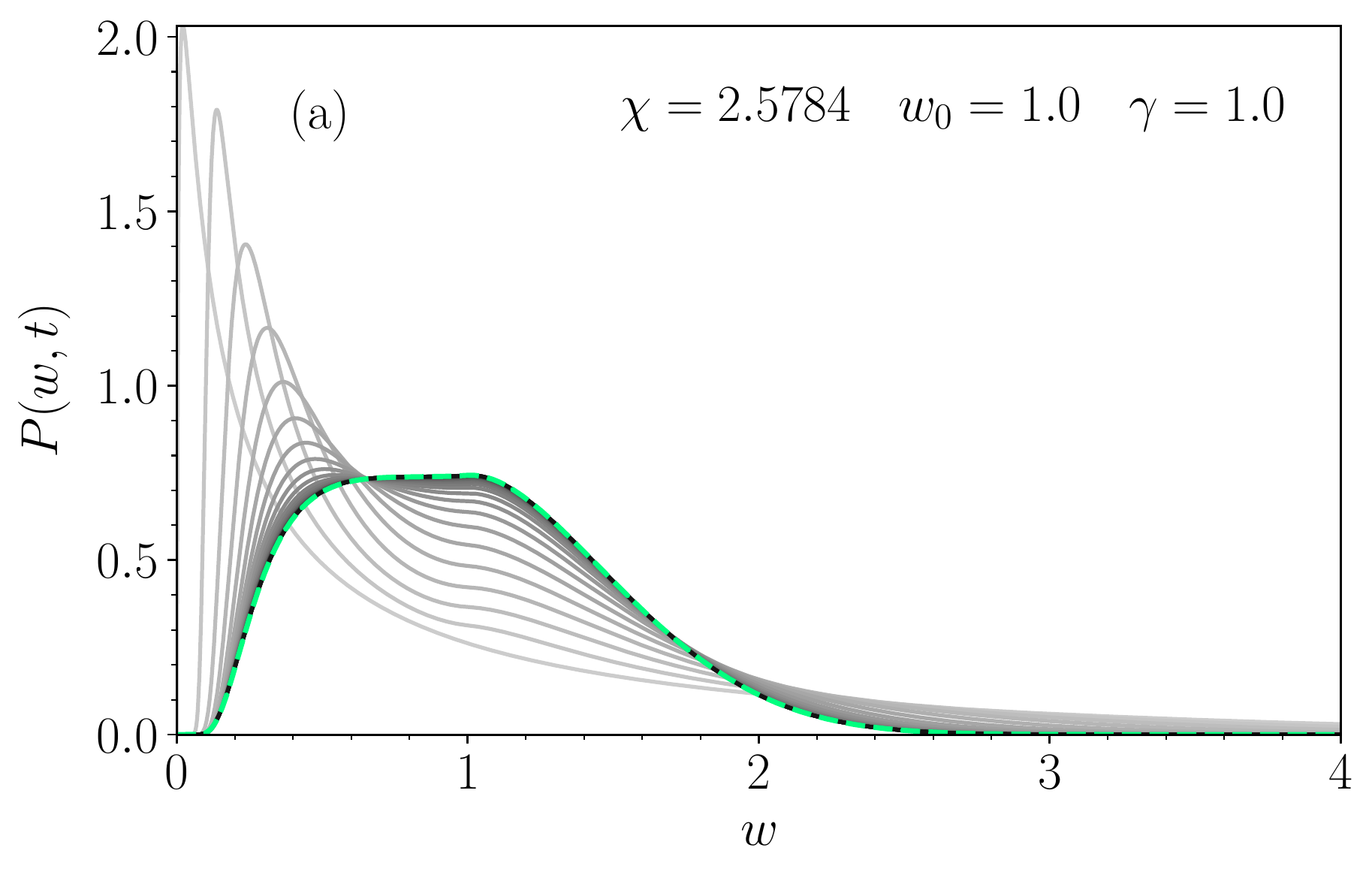}
    \includegraphics[width=0.35\textwidth]{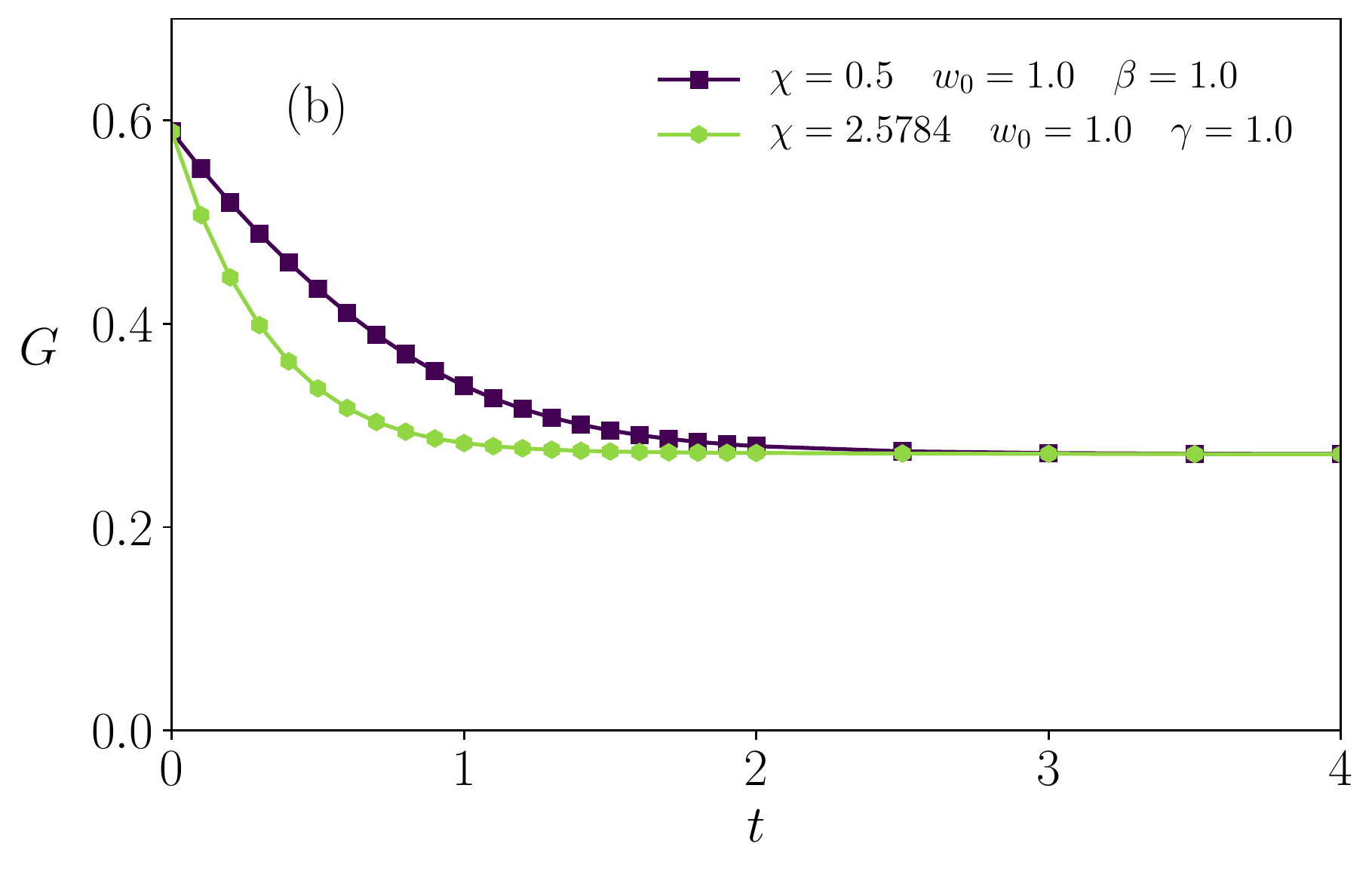}
\caption{ 
	{\bf Setting V (variable subsidies): evolution for piece-wise linear drift.} 
	(a) Wealth PDF (a), $P(w,t)$ vs. $w$, with variable $\beta$, for the same times used in Fig.~\ref{fig:PDFw0_t}. 
	(b) Gini index vs. time for fixed (black) and variable (light green) subsidies, with parameters verifying Eqs.~(\ref{eq:corresp}) leading to the same steady state).
	}
	\label{fig:model2}
\end{figure}

 \begin{figure}[b!]
 	\centering	
	\includegraphics[width=0.379\textwidth]{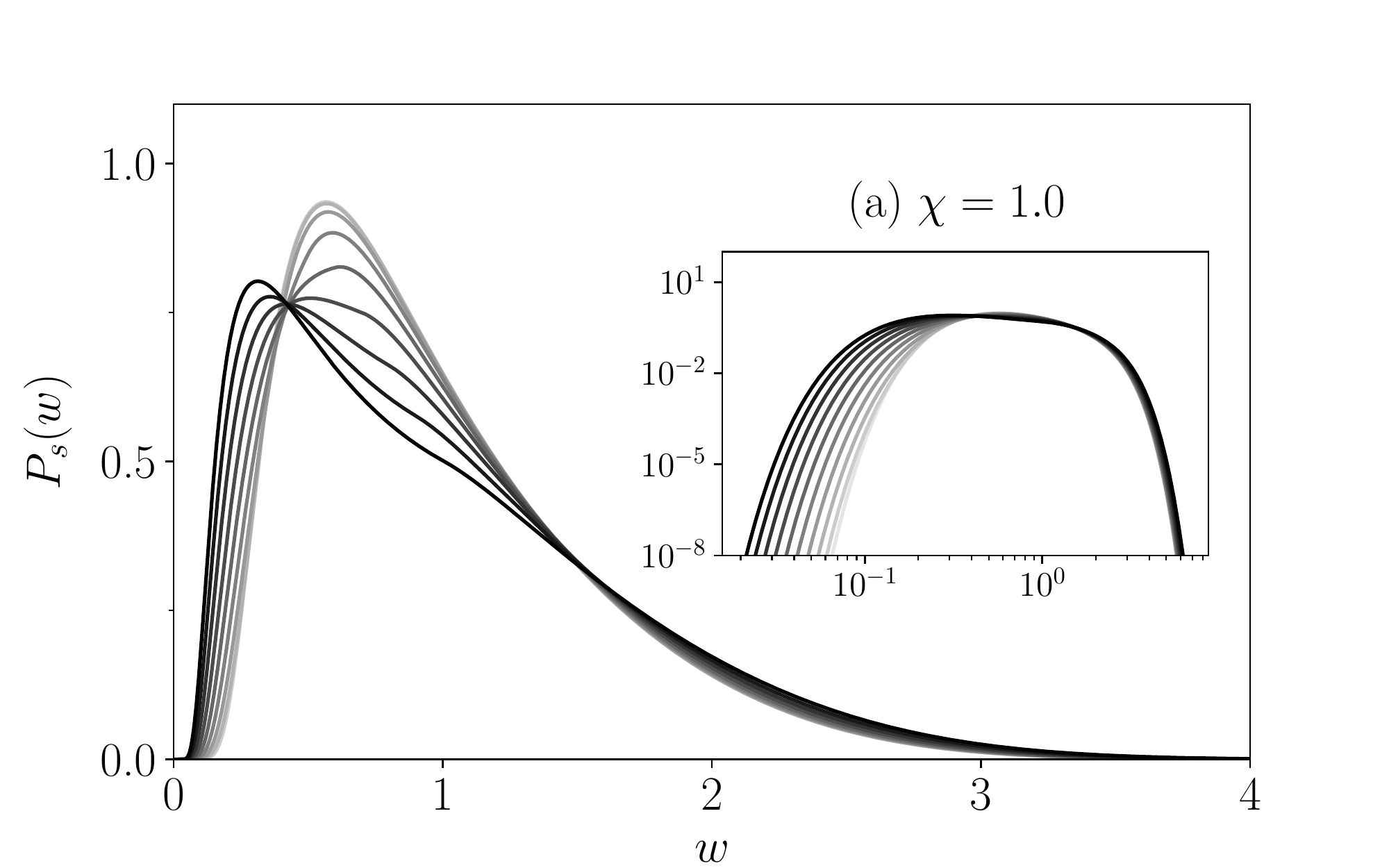}
	\includegraphics[width=0.379\textwidth]{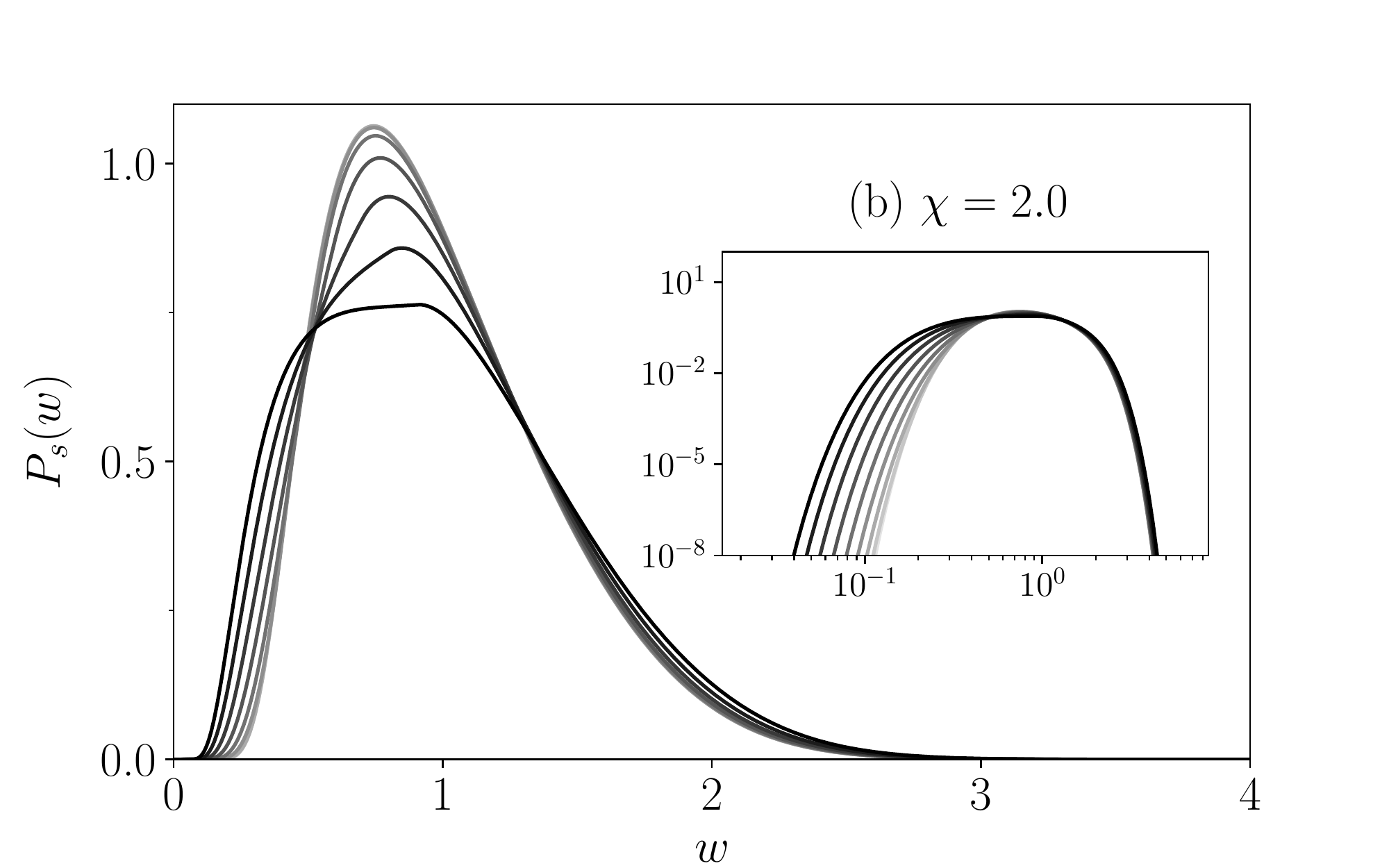}
	\caption{
	{\bf Setting V: stationary wealth PDF with piecewise-linear drift}, for fixed $\chi$ according to the legend, and different values of $w_0$ varying (from light to dark lines) at each  
	$\Delta w_0=0.1$   from 0.0 up to 1.0 in (a) and from 0.0 up to 0.9  in (b). 
	Inset: same data in log-log scale to exhibit the cut-offs.
}
	\label{fig:PDFbeta}
\end{figure}

As an illustration, 
we followed the evolution of a case in scenario F, with parameters 
that verify the above relations with respect to setting V in  Fig.~\ref{fig:PDFw0_t}b. 
In fact, the steady states coincide (compare Figs.~\ref{fig:PDFw0_t}b and \ref{fig:model2}a), however, the temporal evolution differs, as can be also observed by following the Gini indexes over time,  
in  Fig.~\ref{fig:model2}b. The steady state is reached faster
within setting V.

In contrast to setting F, where increasing the threshold $w_0$ reduces the inequality measured by $G_s$, 
in  V, a different dependence on $w_0$ emerges for fixed $\chi$, as shown in Fig.~\ref{fig:model2}b. The stationary value of the Gini index is 
rather insensitive to $w_0$ and even increases with $w_0$. 
A high threshold exempts people with lower level of wealth but the collected wealth is smaller. 
Then it is better to reduce the exemption. 
The corresponding steady PDFs are presented in
Fig.~\ref{fig:PDFbeta}. 
For the power-law taxation, the Gini index decreases with $\alpha$ (not shown). 

Of course, in the limit $w_0\to 0$ or $\alpha\to 1$, yielding the linear case, both setting coincide.

 \begin{figure}[h!]
 	\centering	
	\includegraphics[width=0.379\textwidth]{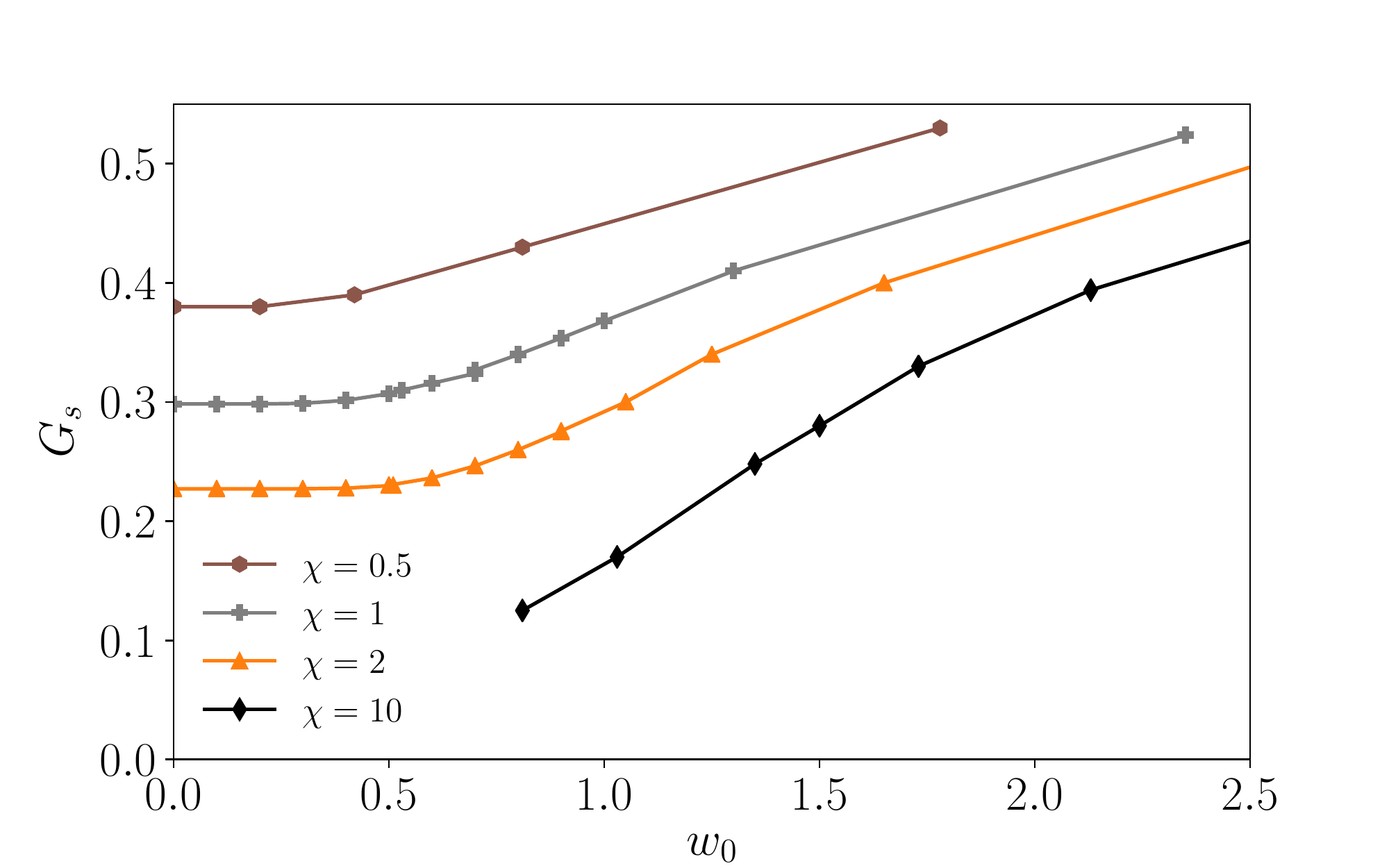}
	\caption{
 {\bf Stationary Gini index for setting V} as a function of $w_0$. Lines are a guide to the eye.    
}
	\label{fig:ginialfaII_s}
\end{figure}

\section{Final remarks}
\label{sec:final}

We studied nonlinear redistributive mechanisms that allow  to reduce inequality. 
Two families of taxation kernels were studied: piecewise linear, with an exempting threshold, and a power-law. 
We compared the results of numerical simulations of the agent-based model with the solutions of the FPE, in very good agreement. 
We performed numerical integration of the FPE  
to follow the time evolution of  $P(w,t)$, starting from a given initial unequal distribution and also the stationary PDF $P_s(w)$ was directly obtained.

The studied tax protocols produce economic mobility against the natural trend 
towards the condensation at $w=0$ that occurs in the regulation-free case. 
The nonlinearity brings new features with respect to the linear case, allowing to achieve greater social equality.
Moreover, distributions with peculiar stylized facts can 
emerge. For moderate values of the control parameters, 
distributions can be bimodal, which indicates 
the stratification and coexistence of the population in  
economic classes with distinct characteristic asset levels. 
A flat profile can also emerge, and, for strong regulation, the skewness of the wealth PDF changes sign. 
 
Furthermore, we discussed the similarities and differences 
between two settings, where either the collected money or the subsidies are fixed.  
Depending on the adjustment made, a high 
exemption threshold  $w_0$ can be detrimental for equality.

Let us also remark that the results can be translated to those of arbitrary $\bar{w}$ by a simple scaling. 
In fact, from Eq.~(\ref{eq:FPE}), we have 
$ P(w,t;\bar{w})= P(w/\bar{w},t;1)/\bar{w}$, together with 
the identifications:  $w_0(1)= w_0(\bar{w})/\bar{w}$ (and $\gamma$ unchanged)  in the piecewise-linear case, and 
$\gamma(1) = \gamma(\bar{w}) \bar{w}^ {\alpha-1}$ in the power-law case.

As a perspective for future work, it would be interesting to include other features such as saving propensity, the possibility of an open economy, as well as spatial structure~\cite{novotny} going beyond the mean-field approximation.

We acknowledge partial financial support by the 
Coordena\c c\~ao de Aperfei\c coamento de Pessoal de N\'{\i}vel Superior
 - Brazil (CAPES) - Finance Code 001. C.A. also acknowledges partial support by 
 Conselho Nacional de Desenvolvimento Cient\'{\i}fico e Tecnol\'ogico (CNPq), 
and Funda\c c\~ao de Amparo \`a Pesquisa do Estado do Rio de Janeiro (FAPERJ).
 We thank Bruce Boghosian for elucidating methods used in his previous works.


\appendix

\section{Numerical integration of the FPE}
\label{app:time}

In order to obtain the solution of Eq.~(\ref{eq:FPE}) 
numerically, we performed the change of variables
\begin{equation} \label{eq:change}
    w = -\ln(1-y),
\end{equation}
that univocally maps the interval $[0,+\infty)$ into 
the interval $[0,1]$~\cite{boghosian2017}.
With this change, the quantities defined 
in Eq.~(\ref{eq:M2}) 
become
\begin{eqnarray}
 A(y,t) &=& \int_y^\infty \frac{P(w(x),t)}{1-x}dx \,, \label{eq:As} \\ 
 B(y,t) &=& \int_0^y  \frac{\ln^2(1-x)}{1-x} P(w(x),t) dx \,, \label{eq:Bs}\\ 
f(y,t)   &=&   \Bigl( \beta\bar{w} - \gamma g(w(y)) \Bigr) \,,  \label{eq:fs}\\ \label{eq:averages}
\bar{w}         &=& -\int_0^1 \frac{\ln(1-y)}{(1-y)}P(w(y),t)dy\,.
\end{eqnarray}
In Eq.~(\ref{eq:fs}), we must consider 
$\gamma=1$ for setting V, and $\beta=1$ for setting F. 
Additionally, the normalization condition, meaning conservation of the population size, becomes
\begin{equation}
    \int_0^1 \frac{P(w(y),t)}{1-y}dy=1.
\end{equation}
  
Then,  Eq.~(\ref{eq:FPE}) can be  rewritten as
\begin{equation} \label{eq:FPEy}
\frac{\partial P}{\partial t} + (1-y)\frac{\partial C}{\partial y} = (1-y)\left\{(1-y)\frac{\partial^2 D}{\partial y^2} -\frac{\partial D}{\partial y}\right\} \,,
\end{equation} 
where  $C$ and $D$ are 

\begin{eqnarray}
 C \equiv C(y,t) &=& \chi f(y,t) P\,,  \\
 D \equiv D(y,t) &=& \frac{1}{2}\left( \ln^2(1-y)\,A+B\right)P  \,.
\end{eqnarray}
 
Equation (\ref{eq:FPEy}) was integrated using a standard 
 forward-time centered-space algorithm.  
  
 \begin{figure}[b!]
	\includegraphics[width=0.45\textwidth]{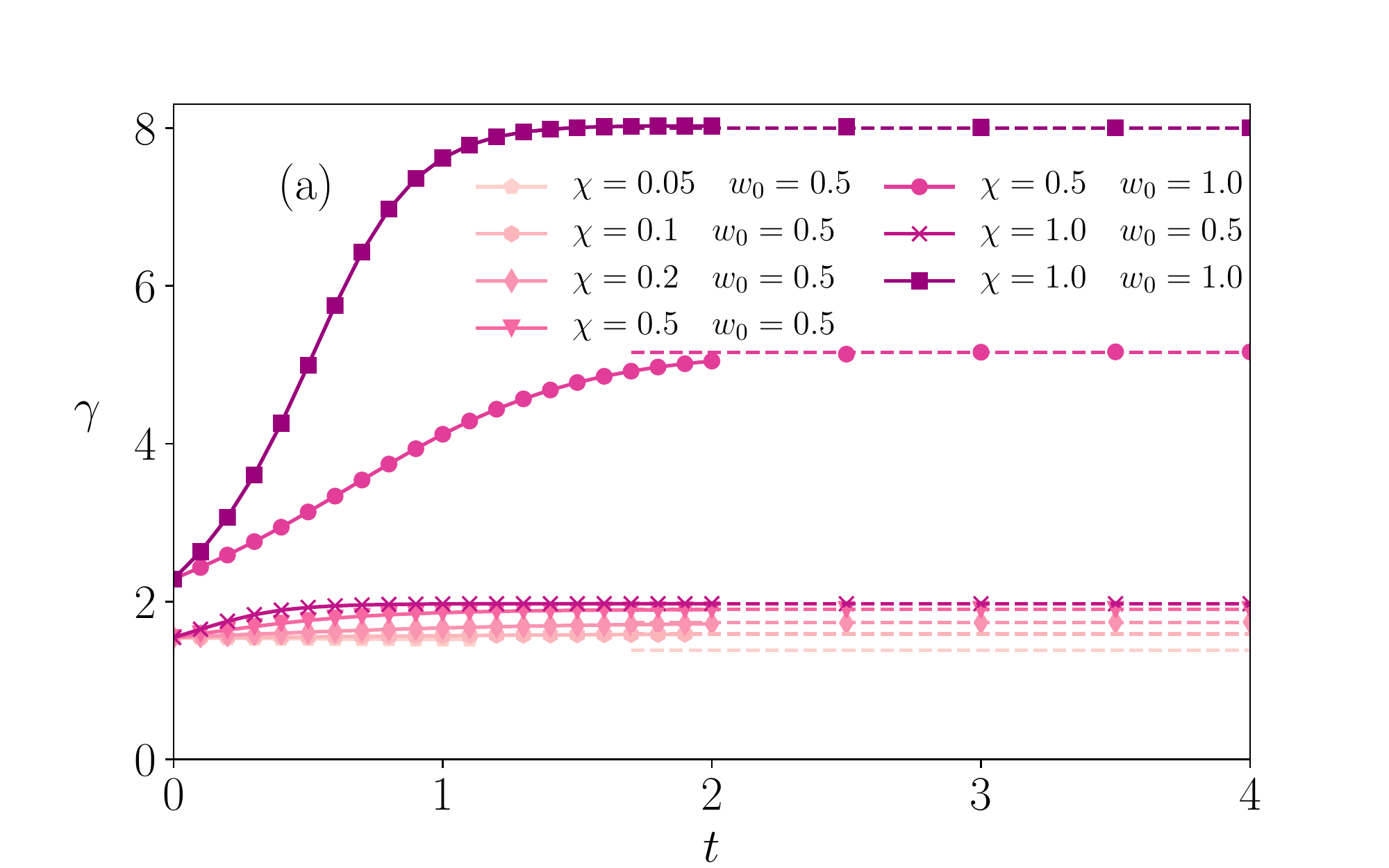}
	\includegraphics[width=0.45\textwidth]{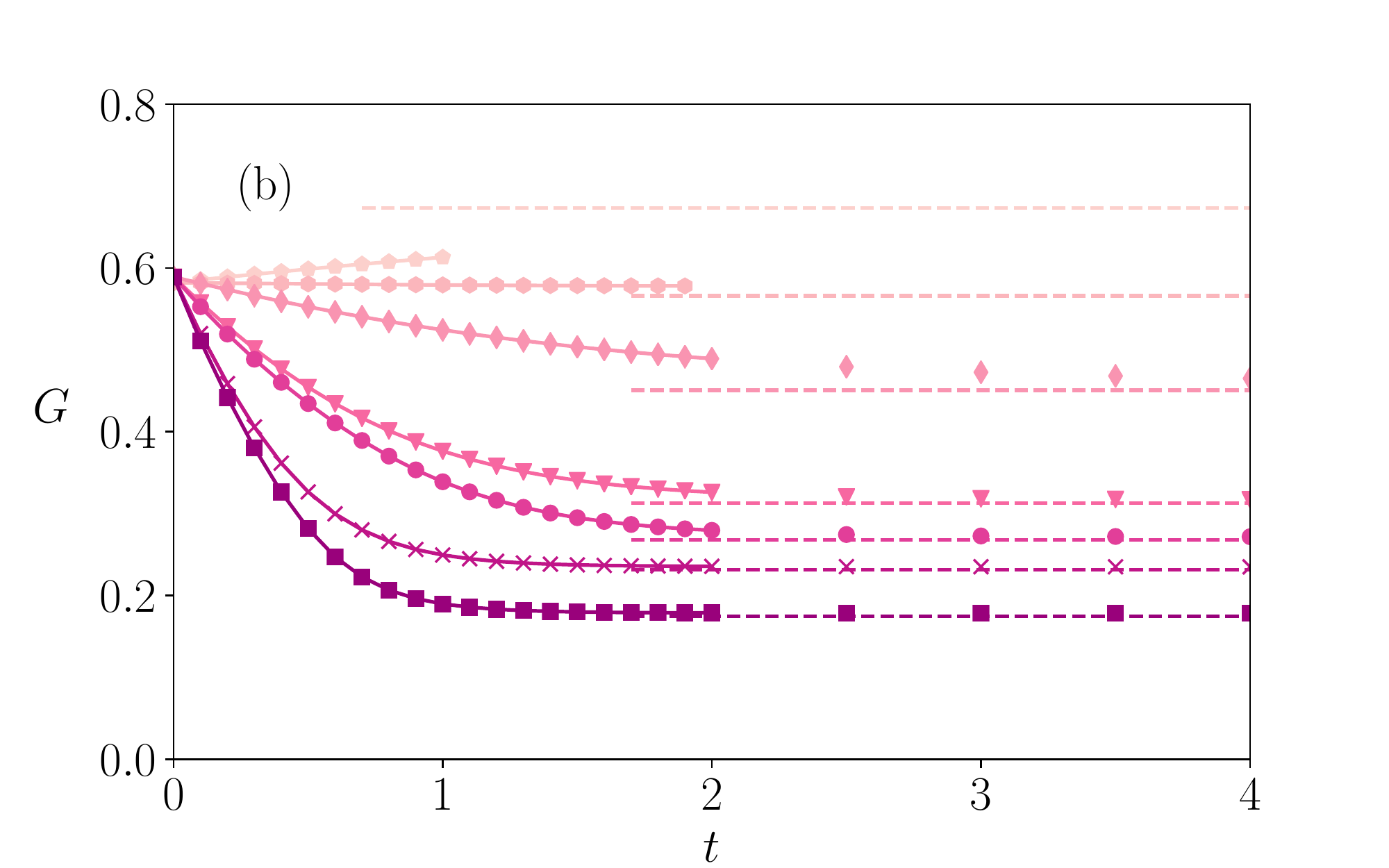}
	\caption{
	(a) Time evolution of $\gamma$, from the numerical integration of FPE (\ref{eq:FPE}), for 
the values of  $\chi$ and $w_0$ indicated in the legend. 
The long-time value of $\gamma$ is in good agreement with that obtained by direct 
integration of the stationary FPE. 
 (b) Evolution of the Gini index $G(t)$.   
Dotted lines indicate the steady state values. 
The computational cost increases when approching the drift-less limit 
$\chi=0$. 
	}
	\label{fig:giniw0_t}
\end{figure}

For the piecewise-linear case, the time evolution of $\gamma$ 
within setting V is shown in Fig.~\ref{fig:giniw0_t}a, 
and that of the Gini index in Fig.~\ref{fig:giniw0_t}b. 
In the absence of regulation ($\chi=0$), condensation would 
proceed, leading to dramatic inequality, with the Gini index monotonically increasing.  
Differently, for $\chi>0$, the Gini index stabilizes. 
Depending on the initial condition, this stabilization can occur from below (e.g., for $\chi=0.05$), or moving far from the condensed state towards  a  more fair distribution of wealth for large enough $\chi$. 
While the initial rate of 
decrease depends on $\chi$, the final stabilization value is ruled by $w_0$ too. %
Of course, for an initial PDF with  sharp cutoff below $w_0$, the drift with be ineffective.


\section{Direct integration of the steady FPE}
\label{app:steady}

In the particular case of the linear rule $g(w)=w$, recovered 
for $w_0=0$ or $\alpha=1$, we have $\beta=\gamma = 1$, independently of the value that $\chi > 0$ assumes, and also independently of time, hence  the stationary value is $\gamma_s=1$ (or $\beta_s=1$).

Otherwise, the stationary value $\gamma_s$ was obtained  
self-consistently 
from the integration of the stationary form of the FPE (\ref{eq:FPE}), which for no flux boundary conditions reads
\begin{equation} \label{eq:FPEs}
\frac{1}{2} \mu ^\prime
-   \chi f P   =0 \,,
\end{equation}
where $\mu\equiv \mu(w) = (w^2 A+B)P(w)$ and ``$\prime$'' mean differentiation with respect to $w$.

In order to solve this equation, we  generalized the numerical procedure described in Ref.~\cite{boghosian2017}. 
It consists in splitting   Eq.~(\ref{eq:FPEs}) into the following coupled linear differential equations, namely, 
 \begin{eqnarray}  \label{eq:set1}
 A^\prime &=&-P =-\mu/(w^2 A+B), \\  \label{eq:set2}
 B^\prime &=& w^2P = w^2 \mu/(w^2 A+B), \\  \label{eq:set3}
 \mu^\prime &=&  2\chi fP = 2\chi f\mu/(w^2 A+B), 
\end{eqnarray}
with the initial conditions $A(0)=1$, $B(0)=\mu(0)=0$, together 
with the normalization condition, namely, $A_0=\lim_{w\to\infty}A(w)=0$.  
This integration however is not straightforward due to the 
singular behavior of $P(w)$ near the origin, that 
behaves as 
\begin{equation} \label{eq:PDF0}
    P(w) \simeq \frac{C}{w^2} \exp\Bigl(2\chi \int^w  \frac{f(x)}{x^2}dx\Bigr)\,,
\end{equation}
where $C$ is a constant. Then, we use the final value in the interval $(0,\delta w)$, with $\delta w \ll 1$ as initial condition 
for the set of differential equations (\ref{eq:set1})-(\ref{eq:set3}).
But  still, this initial condition depends on $C$, which must be  determined from the normalization constraint  
$A_0=\lim_{w\to\infty}A(w)=0$. From the plot of $A_0(C)$ vs. $C$, using a Newton-Raphson (NR) procedure, $C$ can be determined by solving $A(C)=0$, which has a single root. 
This is, essentially, the procedure described before for the 
linear case~\cite{boghosian2017}. In the nonlinear case, 
we must still determine the value of $\gamma_s$ (in setting V) or $\beta_s$ (in setting F) that 
defines $f(w)$, under the constraint of conservation of the  average wealth $\bar{w}$.
Then, a second NR procedure is required to find the correct value of $\gamma_s$ or $\beta_s$ for each value of $C$ that enters in the first NR scheme. 
(Actually in numerical integration, we also use the change of 
variables given by  Eq.~(\ref{eq:change}).)  
The steady solutions  $P_s(w)$ found through this procedure are in accurate agreement with the long-time solutions obtained by numerical integration of the time-dependent FPE, 
as illustrated in the insets of Figs.~\ref{fig:PDFw0_t} and~\ref{fig:PDFbeta}.

\begin{figure}[h!]
\centering
	\includegraphics[width=0.45\textwidth]{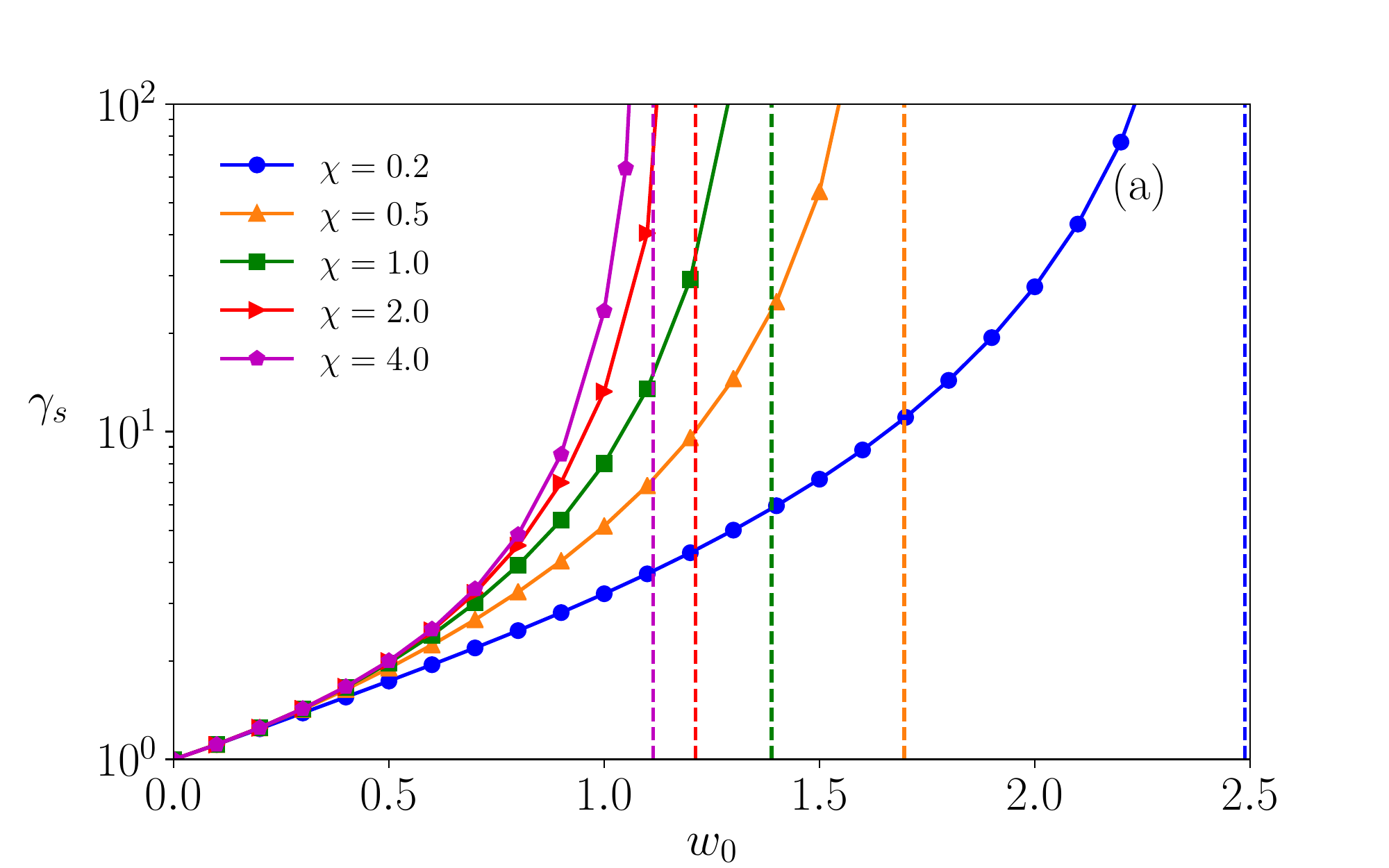}
	\includegraphics[width=0.45\textwidth]{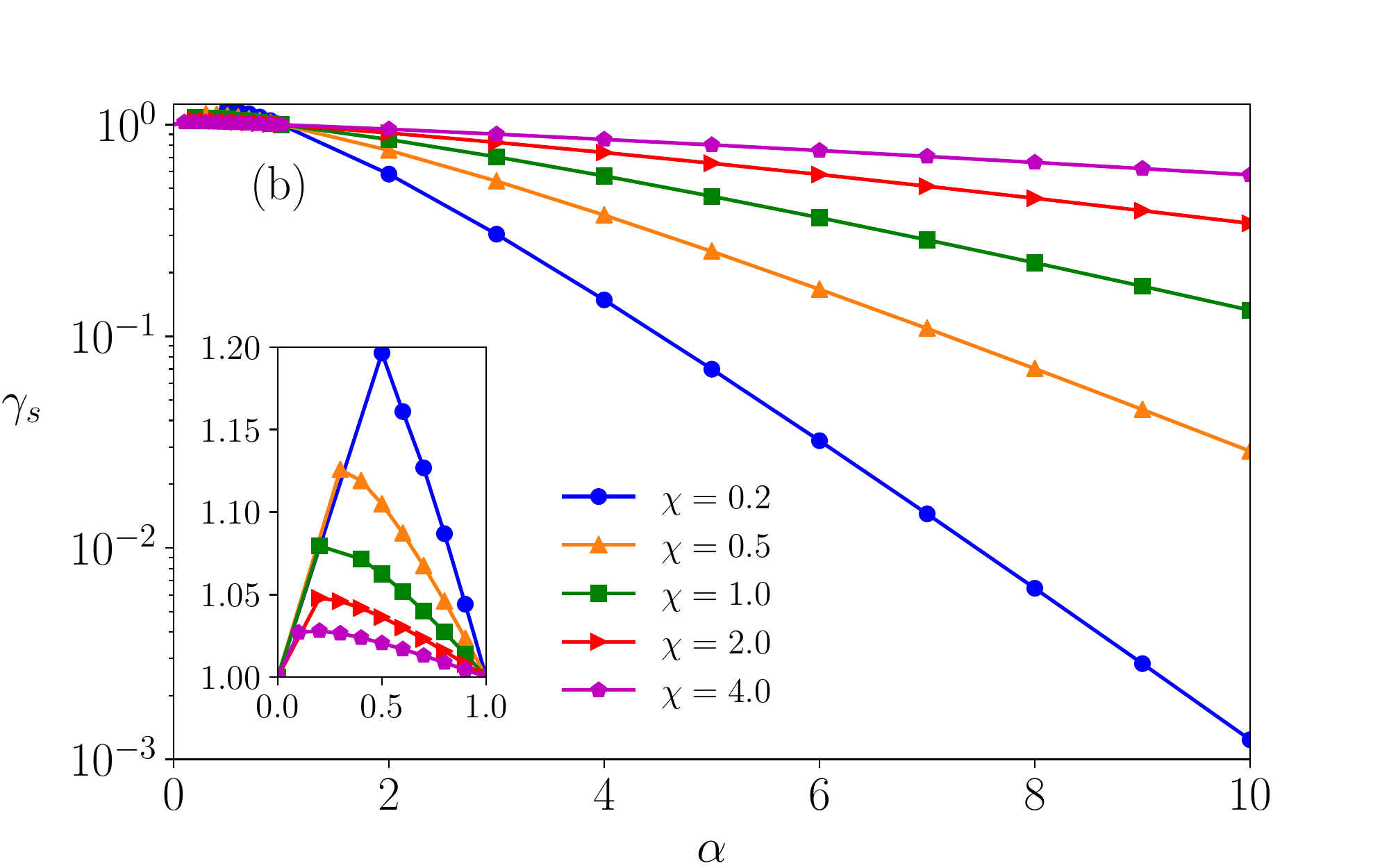}
	\caption{
	Stationary value of $\gamma$ as a function of $w_0$ (a) and  $\alpha$ (b), for different values of $\chi$ indicated in the legend. 
	In (a), the vertical dotted lines indicate the asymptotes. 
	In (b), the inset is a zoom of the vicinity of the origin. Full lines are a guide to the eye. 
	}	  
	\label{fig:gamma_s}
	\end{figure}

For setting F, the stationary value of $\gamma_s$ is 
plotted  in Fig.~\ref{fig:gamma_s}, 
for given values of $\chi$,  as 
a function of $w_0$ (a) and $\alpha$ (b). 
In case (a), $\gamma_s$ increases from unity at 
$w_0=0$ (pure linear case)  diverging at a finite value   $w_{0c}$ (indicated by dotted vertical lines), following the scaling $\gamma_s \sim (w_{0c}-w_0)^{-2}$. 
This is the limit value of $w_0$ observed in Fig.~\ref{fig:giniw0_s}. 
More importantly, it means that although a steady state can be attained, an enormous increase of $\gamma$ would be needed. However, even for intermediate times, at which $\gamma$ assumes moderate values, inequality is reduced.
In agent-based simulations (of finite size), a finite upper bound of $\gamma$ is attained. 

Still in setting $F$, for the power-law rule, the coefficient $\gamma$ 
plays a different role. 
In Fig.~\ref{fig:gamma_s}(b), the curves $\gamma$ vs. $\alpha$ first increase from $\gamma_s=1$,  
up to a maximal value, and then decay exponentially towards zero.
Notice that when, $\alpha=1$ (pure linear case), $\gamma_s=1$. %
When $\alpha=0$, 
  from 
Eq.~(\ref{eq:a0}), we have  $\gamma=\bar{w}$ ($=1$ in our examples). As a consequence the drift term 
with redistributive mechanism vanishes, then we recover 
the dynamics without tax regulation when the system becomes at long time an oligarchy of wealthy people concomitantly with  a condensed phase of the very poor ones  (see inset of Fig.~\ref{fig:gamma_s}b). 
The computational cost increases when reaching the drift-less limit.

\section{Upper and lower bounds}
\label{app:bounds}

The limit values of $w$ can be understood as follows. In the worst case where the subsidy is null, the payment cannot exceed the total wealth, since indebtedness is not permitted in the model. Namely, in simulations it must be
\begin{equation} 
\label{eq:cond1}
     R \,\gamma \,g(w) < w.  
\end{equation}

In the linear case (in which $g(w)=w$ and $\gamma=1$), this implies $R<1$.

For the piecewise-linear kernel, 
Eq.~(\ref{eq:cond1}) implies the condition 
\begin{equation} \label{eq:cond2}
     w< \frac{w_0}{1-1/(\gamma R)}=w_{M}. 
\end{equation}
if $w>w_0$ and $\gamma R>1$, with no restrictions arise otherwise. 
The upper bound $w_{max}$ represents a cut-off, that approaches $w_0$ as $\gamma R$ increases, as can be observed for instance in the PDFs of Fig.~\ref{fig:PDFw0_s}. 
This is in accord with the bounds observed in Figs.~\ref{fig:giniw0_s} and~\ref{fig:gamma_s}(a).

For the power-law kernel, $g(w)=w^\alpha$,  Eq.~(\ref{eq:cond1}) implies the following bounds.

\begin{eqnarray}
    w \le (\gamma R)^\frac{1}{1-\alpha} =w_{M} &&\mbox{$\alpha>1$}\,,\\
    w \ge (\gamma R)^\frac{1}{1-\alpha}=w_{m} &&\mbox{$\alpha<1$} .
\end{eqnarray}

Actually these limits are in excess, since we assumed that the received subsidy is null while it contributes to avoid negative wealth. 
These bounds naturally emerge in simulations, at least for not too large values of $R$ and of the application period $\tau$.


\begin{thebibliography}{99}
\bibitem{footnote} 
We mean individual wealth, in the sense of  amount of money, assets  or goods that  can be eventually transformed into money, owned by each person in a population (independently of being a large amount or not).
\end{thebibliography}
\end{document}